\documentclass[sigconf,authorversion]{acmart}

\AtBeginDocument{%
  }

\copyrightyear{2025}
\acmYear{2025}
\setcopyright{acmlicensed}
\acmConference[ETRA '25]{2025 Symposium on Eye Tracking Research and Applications}{May 26--29, 2025}{Tokyo, Japan}
\acmBooktitle{2025 Symposium on Eye Tracking Research and Applications (ETRA '25), May 26--29, 2025, Tokyo, Japan}
\acmDOI{10.1145/3715669.3725870}
\acmISBN{979-8-4007-1487-0/2025/05}

\usepackage{graphicx}
\usepackage{etoolbox}
\usepackage{caption}
\usepackage{xcolor}
\usepackage{comment}
\usepackage{booktabs}

\usepackage{mathtools}
\usepackage{subfig}
\usepackage{url}

\usepackage{cleveref} 
\usepackage{nicefrac}
\usepackage{multirow}

\newcommand{\makeframegraphics}[3]{%
\subfloat[#1\label{#2}]{%
\frame{\includegraphics[width=.16\textwidth]{#3}}
}
}

\begin{document}

\title[Evaluating Foveated Frame Rate Reduction in Virtual Reality for Head-Mounted Displays]{Evaluating Foveated Frame Rate Reduction in Virtual Reality \\ for Head-Mounted Displays}

\newcommand{\affiliationVISCOM}{%
\affiliation{%
  \department{Visual Computing Group}
  \institution{Ulm University}
  \city{Ulm}
  \country{Germany}}
}

\newcommand{\affiliationVISUS}{%
\affiliation{%
  \department{Visualization Research Center}
  \institution{University of Stuttgart}
  \city{Stuttgart}
  \country{Germany}}
}

\author{Christopher Fl\"oter}
\email{christopher.floeter@uni-ulm.de}
\orcid{0009-0003-8678-7122}
\affiliationVISCOM{}

\author{Sergej Geringer}
\orcid{0000-0001-5147-9785}
\email{sergej.geringer@visus.uni-stuttgart.de}
\affiliationVISUS{}

\author{Guido Reina}
\orcid{0000-0003-4127-1897}
\email{guido.reina@visus.uni-stuttgart.de}
\affiliationVISUS{}

\author{Daniel Weiskopf}
\orcid{0000-0003-1174-1026}
\email{daniel.weiskopf@visus.uni-stuttgart.de}
\affiliationVISUS{}

\author{Timo Ropinski}
\email{timo.ropinski@uni-ulm.de}
\orcid{0000-0002-7857-5512}
\affiliationVISCOM{}

\begin{abstract}
Foveated rendering methods usually reduce spatial resolution in the periphery of the users' view.
However, using foveated rendering to reduce temporal resolution, i.e., rendering frame rate, seems less explored.  
In this work, we present the results of a user study investigating the perceptual effects of foveated temporal resolution reduction, where only the temporal resolution (frame rate) is reduced in the periphery without affecting spatial quality (pixel density).
In particular, we investigated the perception of temporal resolution artifacts caused by reducing the frame rate dependent on the eccentricity of the user's gaze. 
Our user study with 15 participants was conducted in a virtual reality setting using a head-mounted display. 
Our results indicate that 
it was possible to reduce average rendering costs, i.e., the number of rendered pixels, to a large degree 
before participants consistently reported perceiving temporal artifacts.
\end{abstract}



\begin{teaserfigure}
  \frame{\includegraphics[trim={0 3.3cm 0 0.6cm},clip,width=0.59\textwidth]{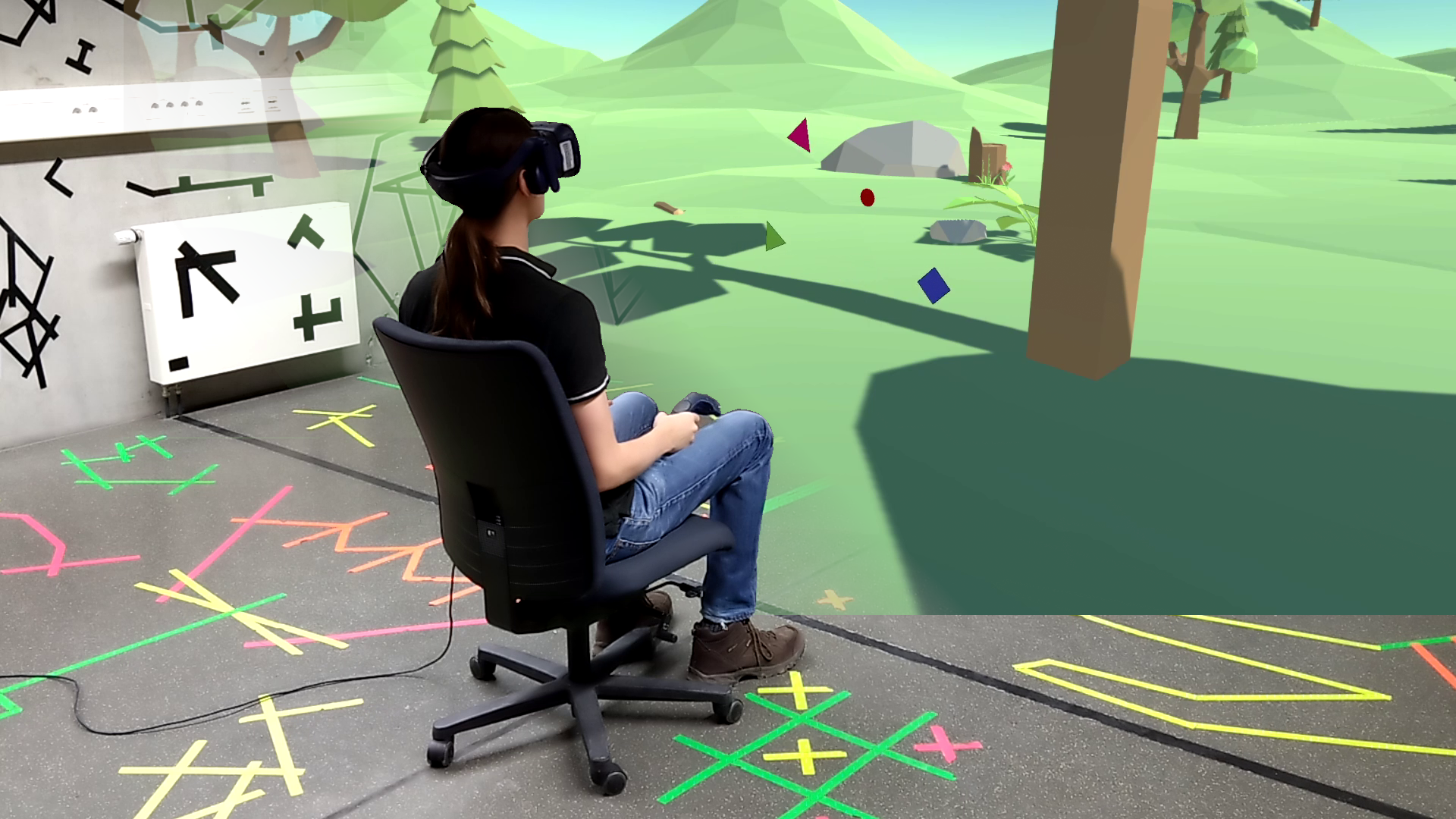}} %
  \frame{\includegraphics[width=0.19\textwidth]{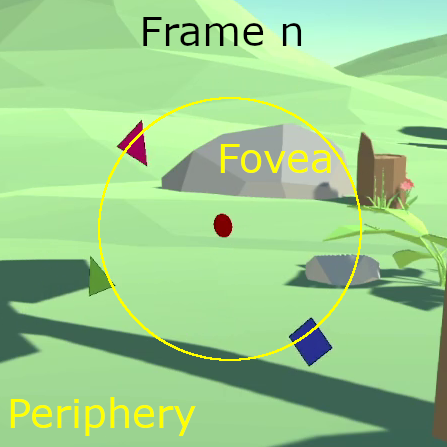}} %
  \frame{\includegraphics[width=0.19\textwidth]{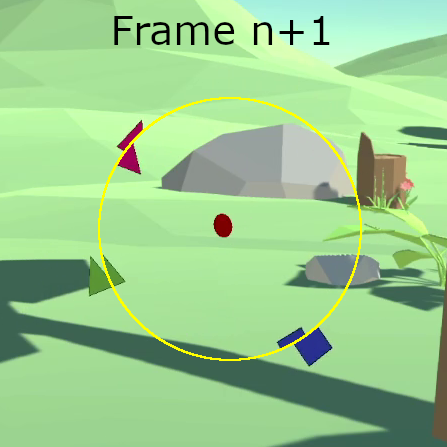}} %
  \captionof{figure}{In our user study, participants tracked objects in virtual reality under different frame rates and scene environments. Frame \textit{n} shows the fully rendered environment, in frame \textit{n}\,+\,1, the periphery is unchanged and only the foveal region is rendered freshly.
  Forest Asset courtesy of Oleh Lila, used with permission.
  }
  \Description{Depiction of our user study setup with three subfigures. The left half shows an artistic image of a user with a VR headset sitting on a chair. In the background, the real environment of the lab is faded to the virtual content the user is experiencing, in this case the forest scene. The right half consists of two figures showing two consecutively rendered frames of the 3D scene, with a circle diving fovea and peripheral zone. In the right frame, objects that cross the circle exhibit tearing as if cut by this dividing circle.}
  \label{fig:teaser}
\end{teaserfigure}

\maketitle

\section{Introduction} \label{intro}

We address the problem of rendering performance for virtual reality (VR) delivered on head-mounted displays (HMDs), with applications ranging from computer games all the way to immersive analytics and visualization. Current HMD technology still falls short of matching human vision, which would require a pixel density of 60 pixels/degree (ppd), a refresh rate of 1800\,Hz, and a field of view (FOV) of $157^\circ\times135^\circ$~\cite{cuervo2018creating}. Today's HMDs, despite their lower resolution, rely on gaze-directed~\cite{levoy_gaze-directed_1990} or foveated rendering~\cite{guenter_foveated_2012} to reduce computational costs and achieve sufficient rendering speed. However, such adaptive rendering is typically restricted to reducing the spatial resolution away from the fovea. In contrast, adapting the temporal resolution is comparatively unexplored~\cite{MOHANTO2022474}, and existing methods for reducing temporal resolution~\cite{Weier2016,dorr2005visibility} are coupled to simultaneously reducing the spatial~resolution.

Our goal is to further advance gaze-adaptive rendering by including foveated temporal resolution reduction, i.e., selectively reducing the image refresh rate (frame rate) in the periphery. 
Therefore, we want to establish a baseline for the perception of temporal artifacts in a virtual reality setting and determine whether it is possible to reduce the temporal resolution in a gaze-dependent way that is not perceivable to the user. 
Our method reduces the temporal resolution in an image region depending on the pixel's distance to the user's gaze.
This defines concentric segments in screen-space around the gaze point that are rendered at different frame rates.
Based on this setup, a user study ($N=15$) was conducted with HMDs to determine the perception of the resulting temporal artifacts and the extent to which they are considered distracting by the user when perceived. \Cref{fig:teaser} illustrates the setup of the user study.
The contributions of this paper are:
\begin{itemize}
    \item We discuss temporal rendering artifacts that occur due to our method of foveated temporal resolution reduction.
    \item We present the results of a user study in VR, reporting subjective user tolerance to the rendering artifacts. 
\end{itemize}
Our results show we can reduce pixel rendering costs by up to $63.6$\% without users feeling uncomfortable. 

\section{Related Work}\label{sec:relatedwork}
\emph{Foveated rendering} is often used to improve rendering performance by reducing rendering quality and workload in the periphery unnoticeable to the user~\cite{MOHANTO2022474}. 
In modern HMDs, gaze information is readily available from built-in eye tracking.
In turn, beyond foveated rendering, virtual- and augmented reality applications use eye tracking to also provide gaze-based user interactions~\cite{Plopski2022, Adhanom2023}.

\citet{MOHANTO2022474} classify foveated rendering methods into four categories: 
(1)~\emph{Adaptive resolution} methods lower the spatial render resolution outside the foveated region. An implementation was presented by \citet{guenter_foveated_2012}. This approach is most commonly used in consumer hardware. 
(2)~\emph{Geometric simplification} reduces the level of detail of objects dependent on gaze. It is often implemented in combination with adaptive resolution rendering.
An implementation is described by \citet{murphy2001gaze}.  
(3)~\emph{Shading simplification and chromatic degradation} reduce the computational effort for calculating the pixel color in a gaze-dependent fashion. 
An implementation is described by \citet{stengel2016adaptive}. 
(4)~\emph{Temporal deterioration} techniques reduce the temporal resolution by reusing already rendered pixels from previous frames, therefore lowering the number of pixels to be rendered.

\emph{Temporal resolution reduction}
has the goal of re-drawing an image, or certain parts of it, at a lower update rate (frame rate) than strictly necessary.
As such, this type of rendering technique need not be strictly gaze-dependent. 
\citet{nehab2006real} presented an approach to re-project (re-use) a fragment when its value has not changed, saving computation costs by delaying pixel updates. 
When perfectly implemented, in theory, there would not be any reduction in render quality. 
Another method by \citet{8643566} decreases the render resolution every second frame. 
The temporal flickering caused by this reduction in quality, given a high enough refresh rate, causes no noticeable loss in quality for users. 
\citet{Weier2016} adapted the sampling density of rays in a ray tracing rendering pipeline based on the eccentricity of the user's gaze. Missing pixels are generated using information from the previous frame in combination with the sample rays of the current frame. 
\citet{dorr2005visibility} defined gaze-dependent image regions and then decreased the temporal resolution by interpolating the currently rendered pixel with the previously displayed value. 
The latter two approaches~\cite{Weier2016,dorr2005visibility} 
facilitate post-processing effects, applying Gaussian blur on foveated regions to reduce the spatial resolution. 
Unfortunately, this makes it difficult to determine the extent to which the temporal artifacts affect the perception of the image quality, since they are always coupled with spatial resolution reduction. 
In this paper, we investigate a more basic rendering strategy compared to previous literature: we re-use pixels from previous frames without applying spatio-temporal re-projection.

\begin{figure}[b]
    \centering%
    \frame{\includegraphics[width=0.49\linewidth]{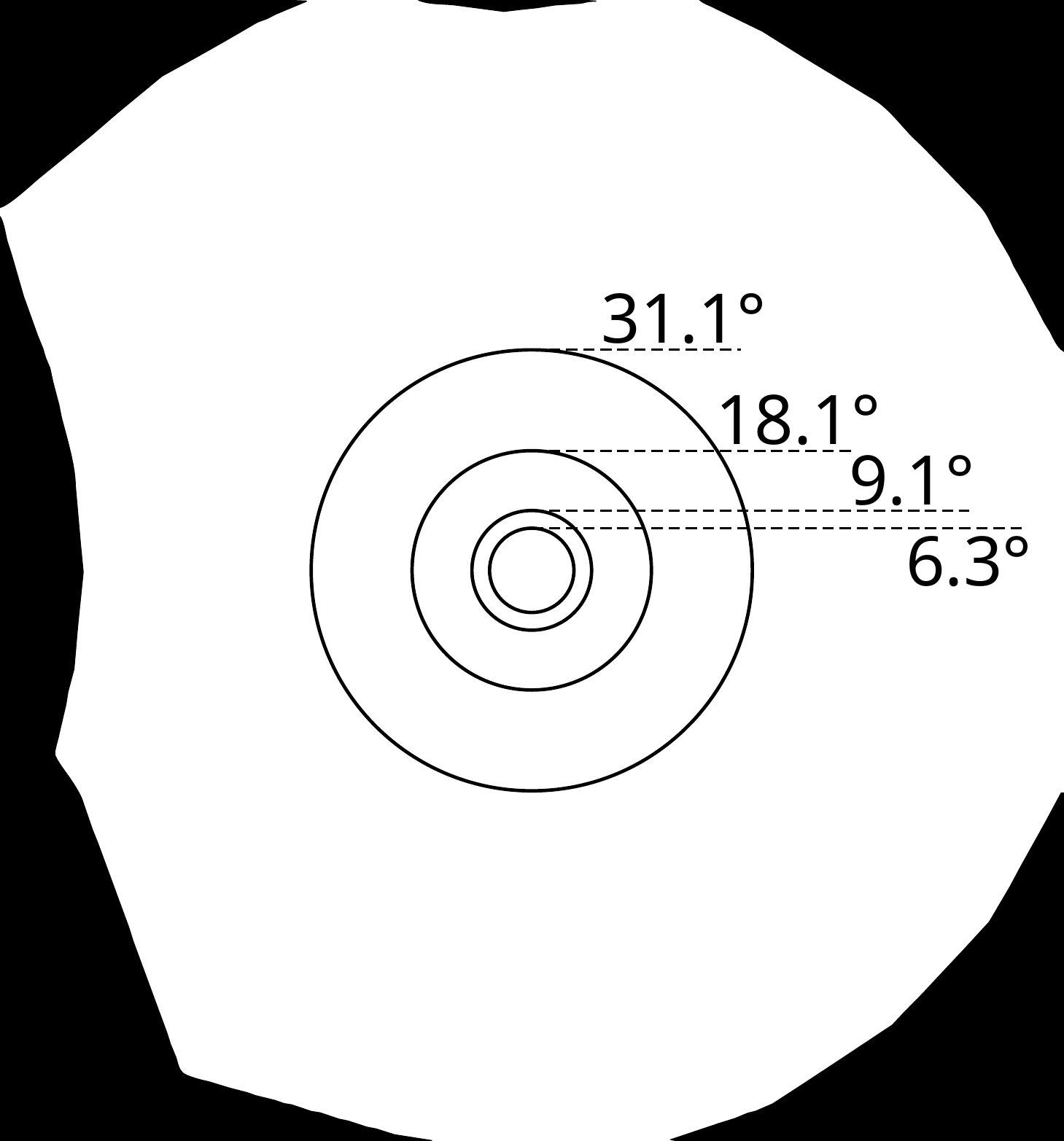}}
    \hfill
    \frame{\includegraphics[width=0.49\linewidth]{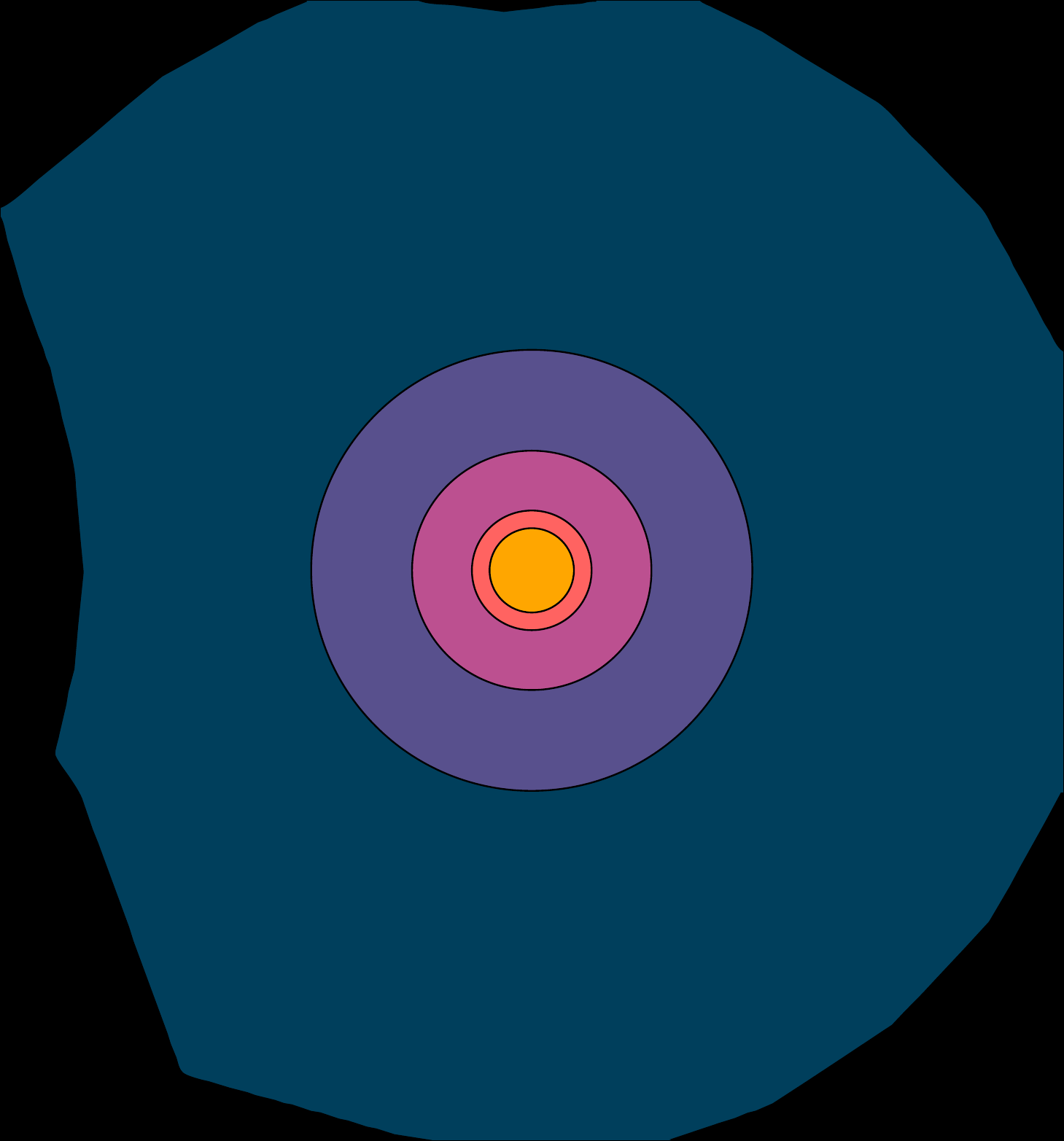}}
     \Description{This figure has two subfigures: The one on the left shows the eccentricity mask, with concentric circles around the center of a silhouette that describes the active pixels for one eye on a VIVE Eye Pro, at angles of 6.3, 9.1, 18.1, and 31.1 degrees, respectively. The image on the right shows a color-coded version, according to frame rate reduction for a 12345 frame rate configuration.}
    \caption{\emph{(Left)} Eccentricity mask defining fovea/periphery image regions, 
    with diameters chosen according to \citet{MOHANTO2022474}.
    \emph{(Right)} Image regions are color-coded according to frame rate reduction for a 12345 frame rate configuration.}
    \label{fig:foveasegments}
\end{figure}

\begin{figure*}[t]
    \centering
    \includegraphics[width=0.95\linewidth]{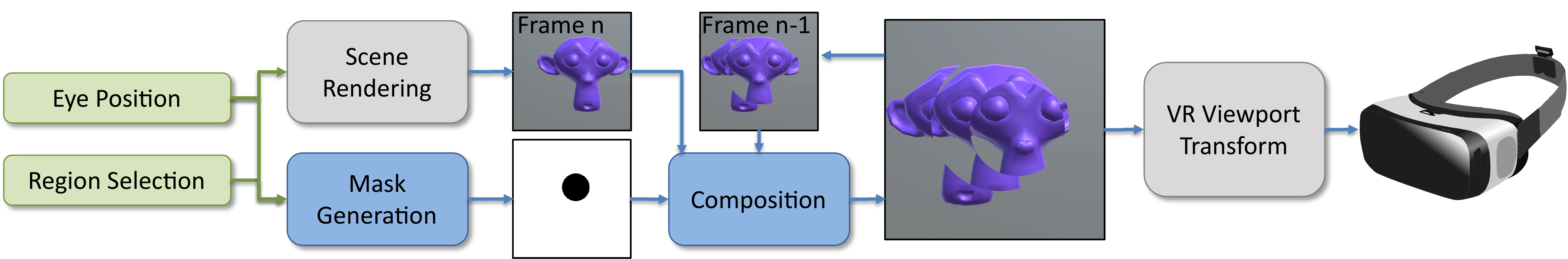}
    \caption{
    Illustration of the rendering pipeline. 
    Green boxes implement foveated rendering, blue boxes implement temporal resolution reduction. 
    The images depict an object moving from the left to the right.
    HMD illustration courtesy of user Juhele ``Jan Helebrant'' via Open Clip Art Library, used under CC0 license. 
    Suzanne Asset courtesy of TurboSquid, used with permission.
    }
    \Description{Illustration of the rendering pipeline. On the left, there are two green boxes that signify the pipeline input: eye position (top) and region selection (bottom). They are both connected to two boxes in a second column: scene rendering at the top, which results in a small representative image of a 3D scene, and mask generation at the bottom, which results in a representative mask image, where black shows updated pixels, and a white background shows static content. The representative image and the mask merge in a composition box. From there, a small loop over the current output image provides the static pixels for the next frame, going back into the composition box. At the right, a VR viewport transform connects the current image to a depiction of a VR headset.}
    \label{fig:pipeline}
\end{figure*}

\section{Method: Technique and Rendering Artifacts}\label{sec:method}

We investigate how reducing the temporal resolution in a gaze-dependent manner is perceived by the user in a VR setting, i.e., we combine the concept of temporal resolution (frame rate) reduction with foveated rendering.

\subsection{Technique: Foveated Temporal Resolution Reduction}\label{sec:technique_ftrr}
\begin{figure}[b]
    \centering%
    \frame{\includegraphics[width=0.49\linewidth]{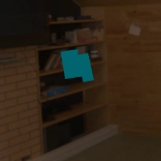}}
    \hfill
    \frame{\includegraphics[width=0.49\linewidth]{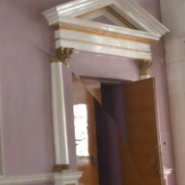}}
    \Description{This figure as two subfigures: The one on the left illustrates temporal image artifacts in the rendered VR image resulting from our rendering technique. The subfigure on the right shows temporal artifacts caused by camera movement.}
    \caption{Temporal artifacts resulting from object movement \emph{(Left)} and camera movement \emph{(Right)} with our rendering method.
    HDRI Room/Attic (left) and Vestibule (right) Assets courtesy of user ``Sergej Majboroda'' via Poly Haven, used under CC0 license.
    }
    \label{fig:tearexample}
\end{figure}

Given the gaze position on the screen, we define concentric circular image segments (foveal/peripheral regions) around the gaze point, depending on the angle spanned by the image pixels and the visual axis, i.e., gaze direction.
We define the image segments using eccentricity angles as provided by \citet{MOHANTO2022474} (see \Cref{fig:foveasegments})%
---resulting in \emph{five} image segments spanning from the fovea to the periphery.
During rendering, we update different image segments at different frame rates: regions in the periphery are updated at slower frame rates than in the fovea. 
\Cref{fig:foveasegments} depicts the image segments based on eccentricity, the rendering pipeline for the method is illustrated in \Cref{fig:pipeline},
and \Cref{fig:tearexample} shows temporal artifacts resulting from this rendering approach.

\subsection{Temporal Rendering Artifacts}
\label{sec:temporal_artifacts}

Updating different image regions at different intervals may introduce temporal rendering artifacts in the form of image discontinuities at region transitions.
This effect is reinforced by our decision to forgo spatio-temporal re-projection, and instead presenting the image regions rendered at different frame rates verbatim to the user.
In the following, we inspect possible causes of these artifacts, and how they might negatively impact the user's experience.
We assume that the VR scene consists of actively moving or stationary objects and a stationary environment (background).

\paragraph{Object and Environment Tearing}
Assuming the user to be stationary, moving objects crossing region boundaries may appear to be cut off. 
Other factors influencing this effect include object \emph{color}, \emph{geometry}, \emph{size}, \emph{movement speed}, and \emph{movement extent}.
Assuming the user moves their head, image tearing of the scene environment (background) will appear across image regions, caused by the non-uniform temporal resolution. 
This effect is determined by the \emph{visual variety} of the environment and the motion induced by \emph{user behavior}.
The described rendering artifacts are exemplified in \Cref{fig:tearexample}. 
The perceived severity of these temporal artifacts depends on the frame rate reduction applied to the individual image regions, i.e., the local rendering delay, and the movement speed of the user and/or objects in the scene.
Thus, the amount of tearing that occurs globally depends on the number of image regions with distinct frame rates.

\begin{figure*}[ht!]
    \centering%
    \makeframegraphics{HDRI street}{fig:sback1}{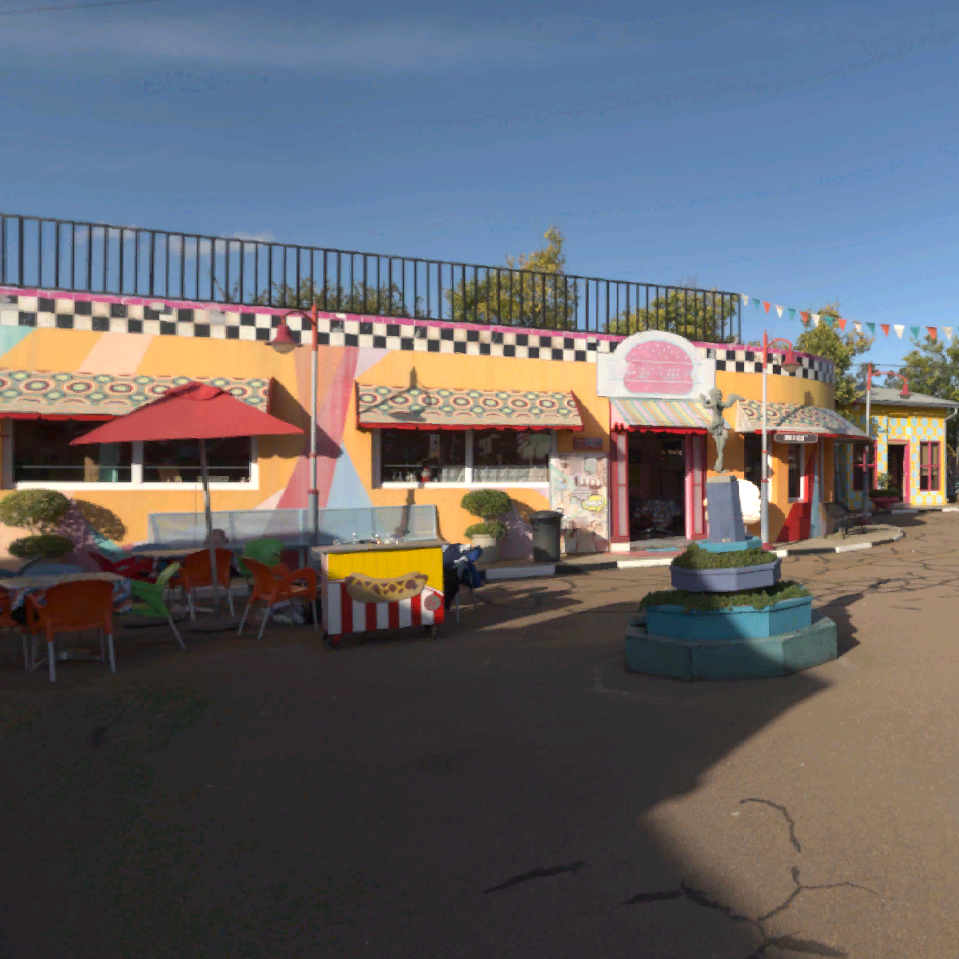}
    \makeframegraphics{HDRI room}{fig:back2}{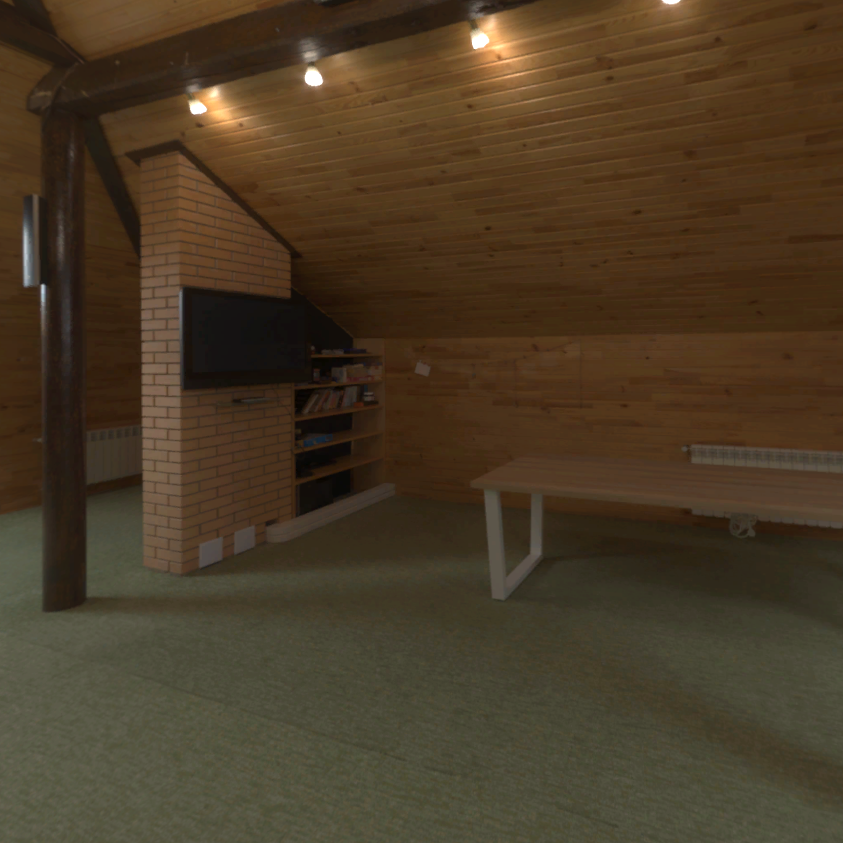}
    \makeframegraphics{Apartment}{fig:back3}{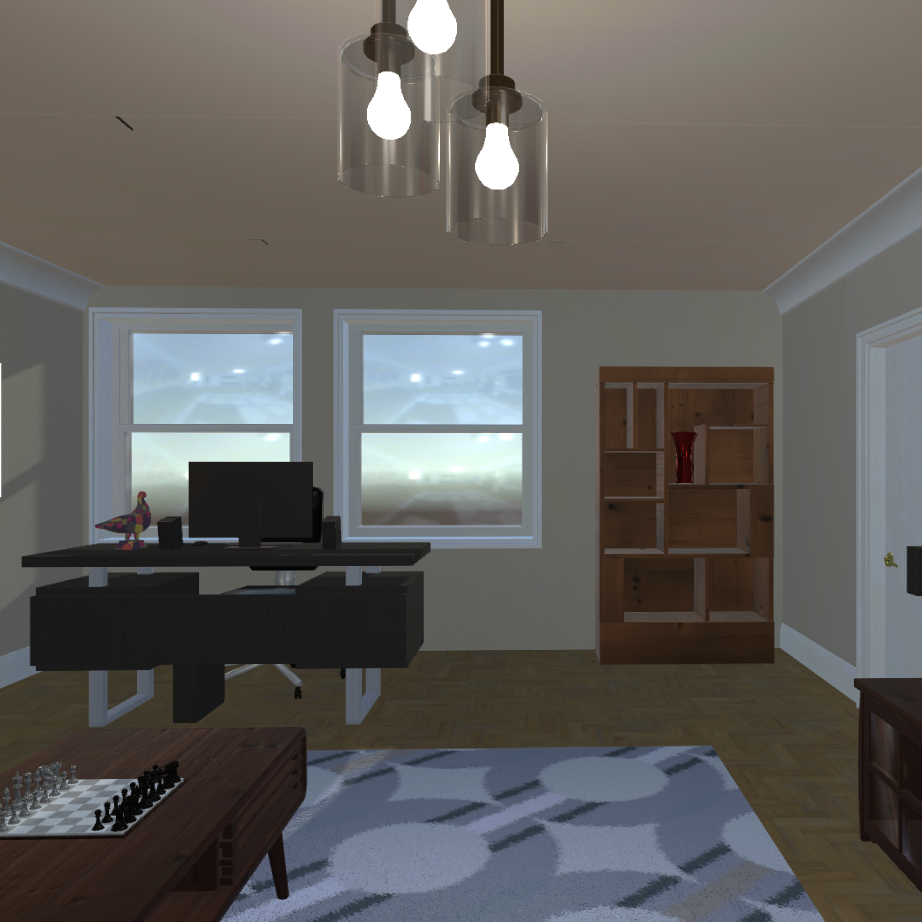}
    \makeframegraphics{Forest}{fig:back4}{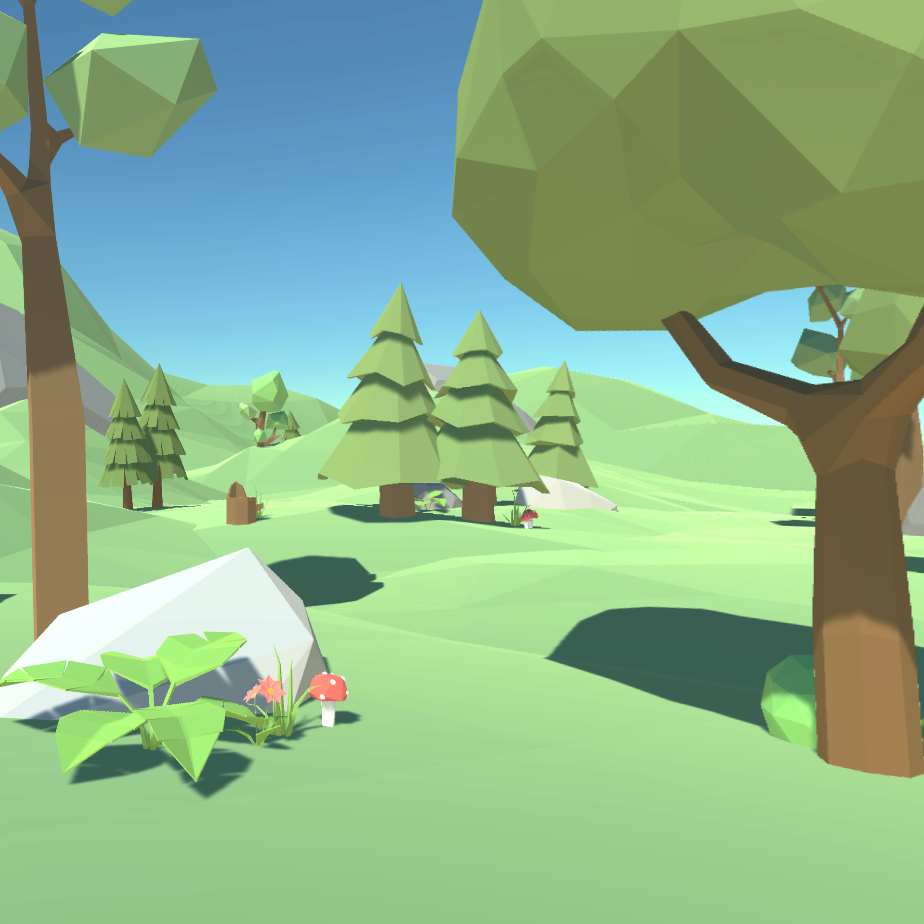}
    \makeframegraphics{Blank canvas}{fig:back5}{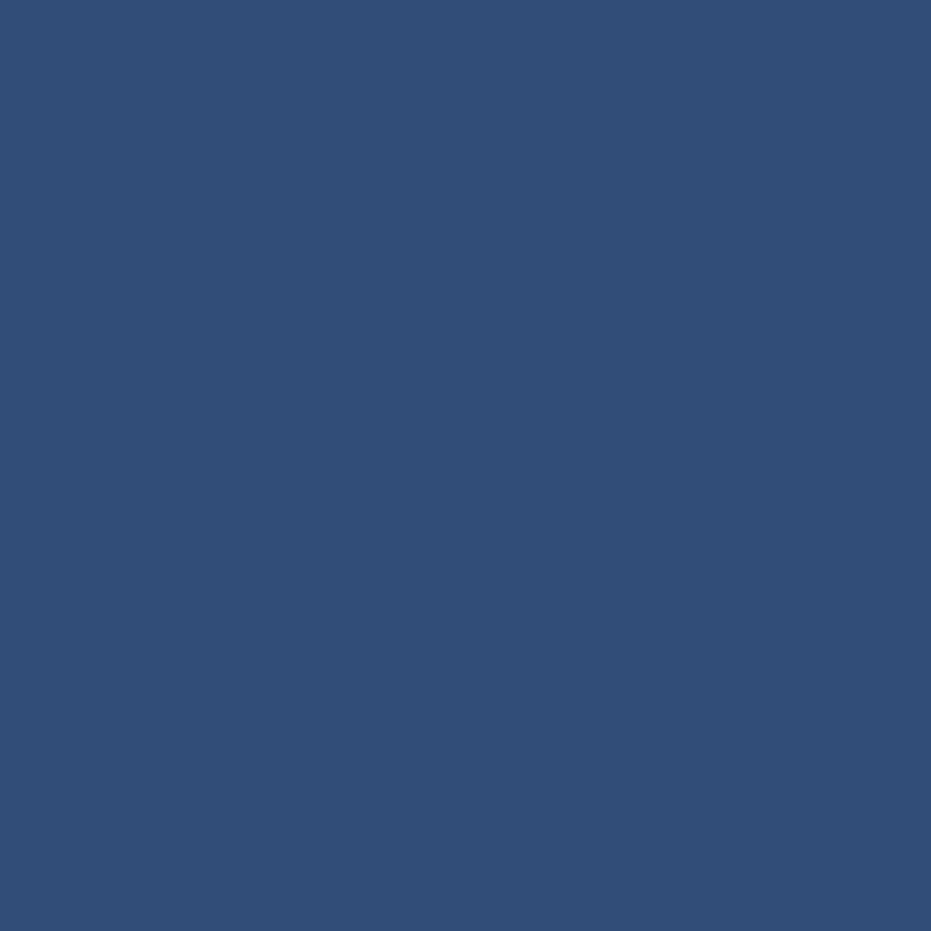}
    \subfloat[Moving objects\label{fig:objects}]{\includegraphics[width=.16\textwidth]{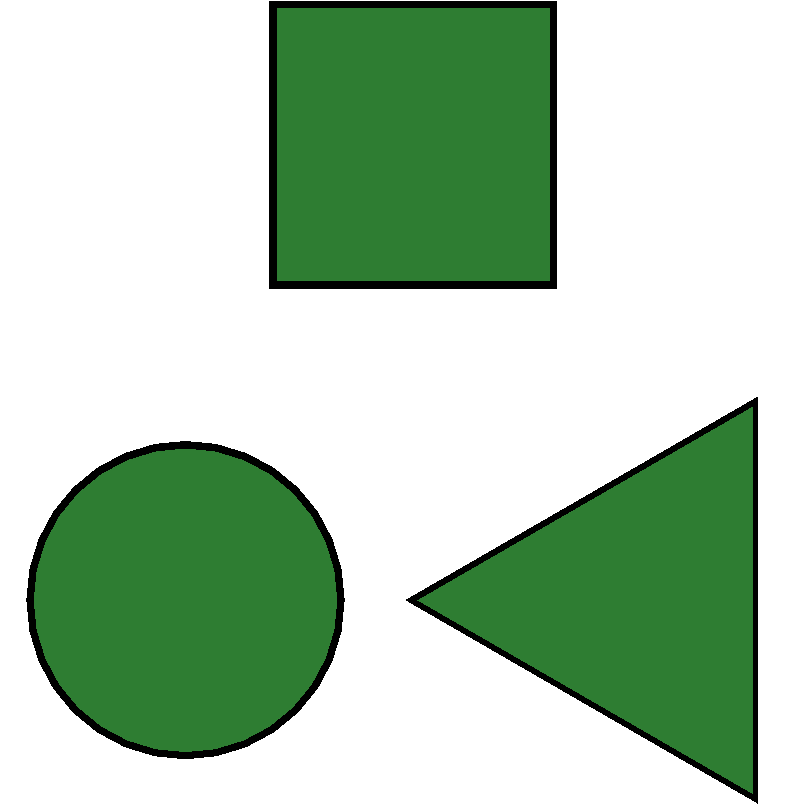}}
    \caption{Screenshots of the environments: \emph{(a, b)} HDRI scans, \emph{(c, d)} pre-rendered scenes converted into cube maps,  and \emph{(e)} cube map with a constant color. \emph{(f)} Object shapes moving in front of the user.
    HDRI Street Asset courtesy of users ``Dimitrios Savva'' and ``Jarod Guest'' via Poly Haven, used under CC0 license.
    HDRI Room/Attic Asset courtesy of user ``Sergej Majboroda'' via Poly Haven,  used under CC0 license.
    Apartment Asset courtesy of Alexis Jose Guevara, 
    used with permission.
    Forest Asset courtesy of Oleh Lila, 
    used with permission.
    }
    \Description{Squared screenshots of the environments used as VR scenes. The images are from left to right: HDRI street scene, HDRI room scene, Apartment scene, Forest scene, and a dark-blue square of a blank canvas. The rightmost image shows small circle, triangle, and square (all green), which serve as object shapes moving in front of the user.}
\label{fig:environments}
\end{figure*}

\paragraph{Image Homogeneity and Scene Motion}
The extent to which the scene affects the perception of the frame rate reduction depends on the image complexity, i.e., the cohesiveness of pixel areas with the same color, which in turn depends on the variety of scene objects and how they are shaded. 
For example, flat or cell shading will lead to larger areas of identically colored pixels than physically-based rendering. 
The latter would lead to more perceived tearing and temporal artifacts.
\Cref{fig:environments} (a)--(e) shows the five scenes we used for our user study, ranging from photorealistic HDRI scans and cell shading all the way to a constant-color skybox. 
The perceived movement of objects, and thus induced tearing, is also influenced by the movement of the camera. 
This is especially the case in a VR environment, where camera movement is immediately driven by the physical head movement of the user.

\begin{table*}[ht!]
\caption{Frame rate configurations (FRCs) and corresponding percentage of pixels that need to be drawn on average per frame.}
\label{tab:reduction_per_configuration}
\begin{center}
\begin{tabular}{c | cccccccccc} 
\toprule
    FRC & 11111 & 11112 & 11122 & 12222 & 11223 & 11233  & 12233 & 12234 & 12334 & 12345 \\
\midrule
    Percentage of pixels & 
$100\%$ & 
$57.6\%$ & 
$52.2\%$ &
$50.2\%$ & 
$36.4\%$ & 
$34.6\%$ & 
$34.4\%$ & 
$27.3\%$ & 
$26.7\%$ & 
$21.5\%$ \\
\bottomrule
\end{tabular}
\end{center}
\end{table*}

\subsection{Implementation}
\label{sec:implementation}

We implemented our rendering method as a VR scene in the Unity engine~\cite{unityengine} version 2022.3, using a post-processing step to combine the latest frame with previously rendered frames according to the fovea/periphery image regions and pre-defined frame rate reduction configurations. 
The implementation we describe in the following also covers the setup and range of possible conditions in our conducted user study.

\paragraph{Scene Setup} 
We implement a scene with a static environment, rendering one of the textures shown in \Cref{fig:environments} as a skybox. 
Further, the scene contains objects with distinct shape and color moving independently and at random in front of the user.
\Cref{fig:teaser} depicts the VR scene with several moving objects in front of the user in a cell-shaded environment.

\paragraph{Temporal Resolution Reduction in Image Regions} %
Our implementation assigns different individual frame rates to different image regions (see \Cref{fig:foveasegments}). 
Starting from the full frame update rate \nicefrac{1}{1}, denoting an image update every frame (in VR usually at 90\,Hz), we define the \emph{frame update rate} \nicefrac{1}{r} as (delayed) image updates every $r$ frames, with $r \in \{1, 2, 3, 4, 5\}$. 
A \emph{frame rate configuration} is defined as a tuple $f = (f_1, f_2, f_3, f_4, f_5) \in \{1, 2, 3, 4, 5\}^5$ 
of
frame update rates (%
\nicefrac{1}{$f_1$}, 
\nicefrac{1}{$f_2$}, 
\nicefrac{1}{$f_3$}, 
\nicefrac{1}{$f_4$}, 
\nicefrac{1}{$f_5$})
for the five image regions. 
Using a shorter notation, the frame rate configuration $11111 \equiv (1, 1, 1, 1, 1)$ denotes frame update rates \nicefrac{1}{1} for all image regions, 
i.e., normal rendering without delays, 
whereas $12345 \equiv (1, 2, 3, 4, 5)$ denotes an update rate of \nicefrac{1}{1} in the central fovea, and decreasing update rates \nicefrac{1}{2}, \nicefrac{1}{3}, \nicefrac{1}{4}, \nicefrac{1}{5} toward the periphery.
For our user study, we consider frame rate configurations with only monotonically decreasing update rates ($f_i \leq f_{i+1}$), where adjacent image regions are updated with at most one frame difference ($f_{i+1} = f_i + 1$  or  $f_{i+1} = f_i + 0$).
To limit the session time per participant 
in the user study we only consider 10 frame rate configurations 
(out of 16 possible configurations according to the above criteria)
: 
11111, 
11112, 
11122, 
11223, 
11233, 
12222, 
12233, 
12234, 
12334, 
12345. 

\paragraph{Metric for Amortized Pixel Rendering Savings} 
To determine the average reduction of pixels rendered per frame for configuration $f$, we define the pixel rendering rate 
$P = \frac{1}{a}  \sum_i p_i  = \frac{1}{a} \sum_i\frac{a_i}{f_i}$.
The amount of pixels rendered on average is $p_i = \frac{a_i}{f_i}$, with $a_i$ being the number of pixels in image region $i$, $f_i$ the frame update rate in that region, and $a$ the total number of pixels on the screen. 
The amortized pixel rendering rates for our frame rate configurations (FRCs) 
are reported in \autoref{tab:reduction_per_configuration}.

\section{User Study}\label{sec:userstudy}

The user study investigates the following research questions regarding foveated temporal resolution reduction:

\begin{itemize}
    \item[RQ1] To what extent are temporal artifacts perceived by the user in a VR setting wearing an HMD?
    \item[RQ2] To what extent does the environment influence the perception of temporal artifacts?
    \item[RQ3] To what extent does the behavior of the user influence the perception of temporal artifacts?
\end{itemize}

\subsection{Study Design, Participants, and Apparatus}
\paragraph{Study Design}
The study was conducted in a controlled lab environment with a within-subject design.
Participants completed tasks under different stimuli in a VR setting while wearing an HMD. 
Study tasks and stimuli were chosen to achieve a total study time of under 60 minutes.
As \emph{independent variables}, we controlled frame rate configurations (10 options) and scene environments (5 options) across two tasks (gaze follow/center task).
As \emph{dependent variables}, we collected subjective user responses on a 5-point Likert scale to questions regarding the perception of rendering artifacts, the grade of perceived distraction, and user discomfort.
Investigating VR sickness~\cite{wang2023effect, tcha2016questionnaire, KIM201866} in our setting would also be desirable but beyond the targeted study duration.

\paragraph{Participants}
We recruited 15 participants from the computer science and medical faculty. 
All participants had normal vision, they reported no visual impairment while wearing the HMD. 
Most participants had very little to no prior experience with VR headsets. 
Participants signed a consent form after being informed about the purpose of the study and their right to withdraw at any time. 

\paragraph{Apparatus}
To ensure interactive rendering rates of 90\,Hz required for a VR setting, a computer with i7-4790 CPU and RTX 2080 GPU was used.
The HMD used was a HTC VIVE Pro Eye with integrated eye tracking. 
With the frame rate configurations discussed previously, the lowest refresh rate for selected image regions was 18\,Hz. 
The VR software for the study was the implementation described in \Cref{sec:implementation}.

\begin{figure*}[ht!]
  \includegraphics[width=.98\linewidth]{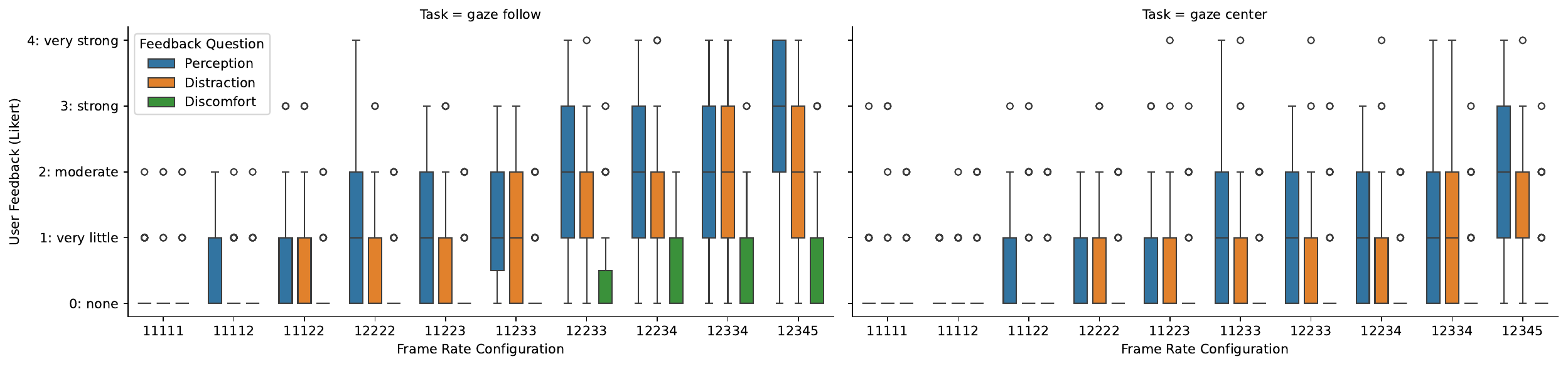}
  \caption{Boxplots showing results of user 
  questions regarding Perception, Distraction, and  Discomfort during the study. 
  The x-axis denotes the tested frame rate configurations, the y-axis denotes user answers on the five-point Likert scale 
  numerically coded from 0 to 4 (merging answers over all scene environments and participants per FRC). 
  \emph{Left}: Gaze follow task. 
  \emph{Right}: Gaze center task.
  }
  \label{fig:results}
  \Description{The figure shows two diagrams with boxplots, representing the results of the feedback questions users answered during the study. 
  The boxplots show user feedback for perception, distraction and discomfort questions on the y-axis on a Likert scale from 0 (none) to 4 (very strong), plotted over the ten frame rate configurations considered in the study setup on the x-axis.
  The left diagram shows boxplot results for the gaze follow task.
  Overall, with increasing rendering delays in the frame rate configurations, the left diagram shows increasingly higher feedback values for the three question categories. 
  The right diagram shows boxplot results for the gaze center task.
  The feedback values in the right diagram also increase with higher rendering delays, but less steep (slower and smaller) than for the left diagram.
  }
\end{figure*}

\subsection{Stimuli and Tasks}

\paragraph{Stimuli}
We employed stimuli consisting of a combination of the ten frame rate configurations and five scene environment settings, along with objects moving in front of the user (see \Cref{sec:implementation}).
Each stimulus was presented to a participant once in each of the two tasks, for a total of $10 \times 5 \times 2 = 100$ stimuli (study rounds).
For the definition of the fovea-centered image regions, $1.1^\circ$ was added to the size of the original regions to accommodate for the tolerance in gaze determination of the used HMD. The number and size of the image regions were constant for the entire trial. 
The three moving objects were changed in color and geometry every round. 
During Task~1, all three objects had the same random color and geometry.
For Task~2, the color and geometry of all objects were independent and random.
During the study, the object speed and movement area were fixed for each round because increasing the speed of the objects, or the extent of their movement, would have changed the temporal artifacts in different image regions.

\paragraph{Tasks} 
We employed two different tasks 
based on the factors influencing temporal rendering artifacts discussed in \Cref{sec:temporal_artifacts}.
Prior research suggested that the perception of temporal artifacts depends on whether the user is tracking an object or focusing on a stationary point~\cite{Weier2016}.
The \emph{gaze follow task} (Task~1) consisted of three moving objects of the same shape and color. These objects moved independently, at random, in a specified region for 15 seconds. At the beginning of the round, one object was highlighted for two seconds. At the end of the round, the participant was asked to select the object highlighted in the beginning. This was done to encourage the user to follow a moving object.
The \emph{gaze center task} (Task~2) showed a small stationary circle for the entirety of the round. At the beginning of the round, the circle was green and turned red after 5 to 10 seconds. Once the circle turned red, the participant was asked to react as fast a possible by pressing a button on a controller. To ensure the presence of temporal artifacts, the moving objects from the first task were also present, whereby each of the three objects had a random shape and color.

\subsection{Procedure}
Participants were introduced to the VR controls and calibrated the HMD eye tracker.
Participants then performed four test trials without temporal resolution reduction to get familiar with the VR scene and to get a frame of reference on the quality of the VR environment. 
For the duration of the trial, the participants were stationary and seated.

During the study, 
participants carried out the tasks under the stimuli described above, 
performing one task after the other with a randomized order of its stimuli.
Learning effects were compensated by counter-balancing the order of the tasks.
After every round, the participants provided feedback 
in the virtual environment, 
answering questions 
 whether they could perceive any visual artifacts (FQ1 Perception), 
 if the perceived artifacts were distracting (FQ2 Distraction), and
 if they were feeling any discomfort (FQ3 Discomfort), 
rating their subjective experience about the round on a 5-point Likert scale (``none,'' ``very little,'' ``moderate,'' ``strong,'' and ``very strong'').
Apart from answering the feedback questionnaire, participants had no breaks between stimuli or tasks.

\section{Results}\label{sec:results}
\Cref{fig:results} summarizes answers for the three measured dependent variables across both tasks.
For the gaze follow task, discomfort stays low, for the gaze center task it is even negligible (green boxes).
Over all tasks and trials, discomfort values of two and above (``moderate'' or worse) were rare except for one specific participant.
This is reflected in the outliers for the discomfort answers.
Perception (blue boxes) and distraction (orange boxes) variables increase more quickly for the gaze follow task than for the gaze center task.
Looking at the measured frame rate configurations more closely, the baseline FRC 11111, apart from outliers, predominantly shows the lowest values (``none'') for the perception, distraction, and discomfort variables across both tasks.
For the gaze center task, this also applies to FRC 11112.

For the gaze follow task, the upper quartile of \emph{perception} rises to ``very little'' as soon as the periphery gets less than full update rates, and becomes ``moderate'' as soon as at least three regions have reduced update rates.
The upper quartile of perception only becomes ``very strong'' when the worst configuration is used (FRC 12345).
For the gaze center task, the perception upper quartile is only ``strong'' for FRC 12345, the four preceding configurations still are ``moderate'' (only the best of these is ``moderate'' for the gaze follow task).
The upper quartiles of \emph{distraction} rise more slowly, capping at two ``strong'' configurations for the gaze follow task, and two ``moderate'' configurations for the gaze center task.
The gaze center task then has ``very little'' distraction down to all four outer regions with halved update rates.
The gaze follow task needs the two outermost regions to have halved update rates for distraction to rise to ``very little,'' followed by three ``moderate'' configurations.
\emph{Discomfort} begins to be noticeable in the gaze follow task, starting from FRC 12233.
Even with the worst configuration, FRC 12345, the upper quartile for discomfort still only reaches ``very little.''

User feedback grouped by the five \emph{scene environments} is very similar for identical frame rate configurations.
The blank background seems to cause slightly lower (better) scores across all frame rates, but only for the gaze center task.
Somewhat unexpectedly, for the gaze follow task, results for the blank background are not better than for other environments.
We provide additional plots detailing these aspects, as well as others, in the supplemental material.

\section{Discussion} \label{sec:discussion}
We investigated our first research question, (RQ1)  perception of temporal artifacts by the user (see \Cref{sec:userstudy}), by collecting user feedback across the dimensions perception of artifacts (FQ1), their distraction (FQ2), and user discomfort (FQ3).
With our chosen frame rate configurations, we wanted to cover a large range of configurations to allow detailed analysis of user sensitivity to frame rate reduction, without having to invest too many resources into strictly finding just-noticeable differences (JNDs).

The results presented in \Cref{sec:results} show that for both tasks, (1)~users increasingly perceive frame rate reduction with smaller frame update rates and (2)~frame rate reduction is perceived as more distracting or uncomfortable with smaller update rates toward the foveal region. 
For example, shifting from configuration $11223$ to $11233$ (one additional update rate $3$ in the periphery)
still shows similar results across all three feedback dimensions, with the latter configuration being rated slightly worse. 
At the same time, configurations $12222$ and  $11223$ are rated similarly, although the former uses an update rate $2$ in all regions except for the most central, whereas the latter shows the ground truth (update rate $1$) in a larger foveal region and an update rate of $3$ in the outermost periphery.
However, for all of these three FRCs, less than 50\% of users report no or barely perceivable artifacts and, at the same time, not being distracted or feeling uncomfortable (see median markers in the boxplots).

Regarding the influence of user behavior on the perception of temporal artifacts (RQ3), results of the \emph{gaze follow} and \emph{gaze center} user tasks show a clear tendency for users to perceive artifacts more, being distracted by them or even feeling uncomfortable, when they have to actively follow objects. 
A little bit surprising for us is the result from the gaze follow task that even for small frame update rates, e.g., configuration $12345$, although users perceive artifacts and are moderately distracted by them, they still report low discomfort scores mostly in the range of ``none'' or ``very little'' discomfort. 
%
None of the participants had to interrupt or abort the trial due to VR sickness. 
Two of the participants experienced moderate symptoms of VR sickness, one of them reported they knew they were prone to it, the other had no prior VR experience. 
Other forms of discomfort reported by participants were due to the HMD feeling uncomfortable and eye strain due to focusing over an extended period of time. 
For the gaze center task, 
users mostly reported low discomfort across all FRCs. 
This could be attributed to our rendering method producing temporal artifacts only after large head movements by the user, which were discouraged in the second task by design.

Regarding influence of the scene environment (RQ2), we found no significant differences in user feedback for frame rate configurations when grouped by scene environment. 
Surprisingly, the monotone-colored scene background was rated similar to the other environments across both tasks.

Literature suggests that VR sickness levels increase 
when exposed to a VR environment over a long period of time~\cite{moss2011characteristics, doi:10.1518/hfes.45.3.504.27254}.
Thus, we assumed that the VR setting and our stimuli would cause increasing discomfort over the study length.
We found that about half of the participants did not experience significant discomfort at all. Only two participants showed a clear increase in discomfort over time. 
We need further studies to ascertain whether there is a dependency or not. 
For example, the length of our stimuli (about 10--15 seconds) may be too short, and longer exposure to frame rate reduction, e.g., 1--2 minutes or more, may show greater discomfort or sickness effects on participants.
There was no measurable learning effect observed between the different tasks (see the supplemental material).

\section{Ethical Considerations and Privacy}
General ethical and privacy concerns related to measuring eye tracking data apply to this work as well. Because foveated rendering 
does not need to store such data privacy issues can be mitigated.

\section{Conclusion and Future Work}\label{sec:conclusion}
In this paper, we evaluated foveated frame rate reduction, a foveated rendering method where the frame update rate is varied in different regions of the rendered image. 
The results of our user study show that it is possible to substantially reduce the amount of rendered pixels in head-mounted VR before users consistently notice the altered frame rate updates and report discomfort. In our specific setting, a reduction by $63.6\%$ (FRC $11223$) is feasible. 
This observation is encouraging, indicating the potential of temporal resolution reduction. 

However, further research is required as the intuitive judgment of our proposed technique would expect more and more severe discomfort and VR sickness feedback by users---while most of our participants reported barely feeling discomfort or being distracted by the rendering technique. In particular, extended user studies could include more participants and a wider range of stimuli and tasks. Another direction of future research could integrate spatio-temporal re-projection into our rendering method.

\begin{acks}
The authors wish to thank the anonymous reviewers for their constructive and critical feedback.

We also thank the credited asset authors and organizations for granting permission to use their works for publication in this paper.

This work was partially funded by the 
\grantsponsor{DFG}{Deutsche Forschungsgemeinschaft (DFG, German Research Foundation)}{https://www.dfg.de}---Project-ID \grantnum{DFG}{251654672}---TRR 161, 
and the 
\grantsponsor{MWK}{Ministerium f\"ur Wissenschaft, For\-schung und Kunst (MWK, Ministry of Science, Research and Arts) Baden-W\"urttemberg, Germany}{https://mwk.baden-wuerttemberg.de}---Project \grantnum{MWK}{Virtuelle Kollaborationslabore Baden-W\"urtemberg, Phase 1--3}---KoLabBW.
\end{acks}

\bibliographystyle{ACM-Reference-Format}


\clearpage
\appendix


\begin{figure*}[hb!]
\begin{tabular}{p{0.48\linewidth} p{0.48\linewidth}}
{
\section{Supplemental Material}
\label{sec:supplemental}
This supplemental material provides further details regarding the evaluation for our paper ``Evaluating Foveated Frame Rate Reduction in Virtual Reality for Head-Mounted Displays''.
The original paper is available at: \url{https://doi.org/10.1145/3715669.3725870}
} & \multirow{2}{*}{
  \includegraphics[width=.43\textwidth]{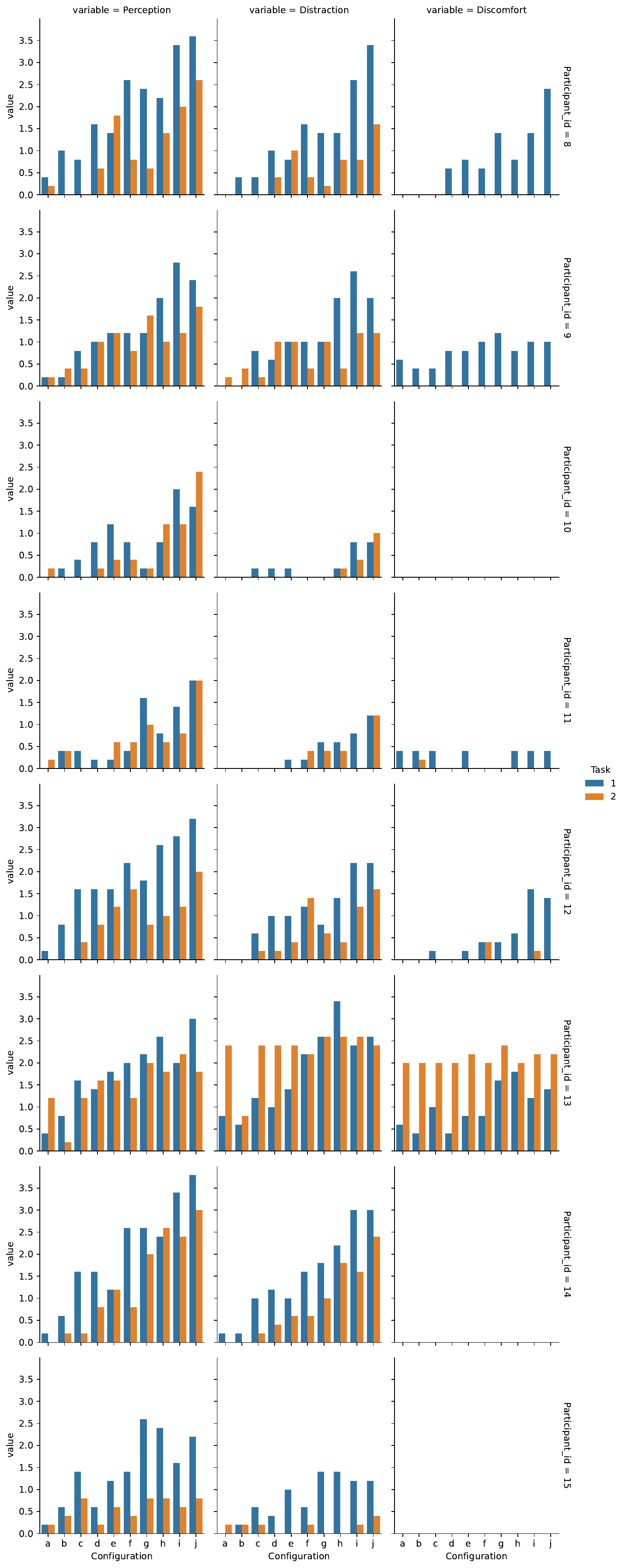}  
}\\
\includegraphics[width=.43\textwidth]{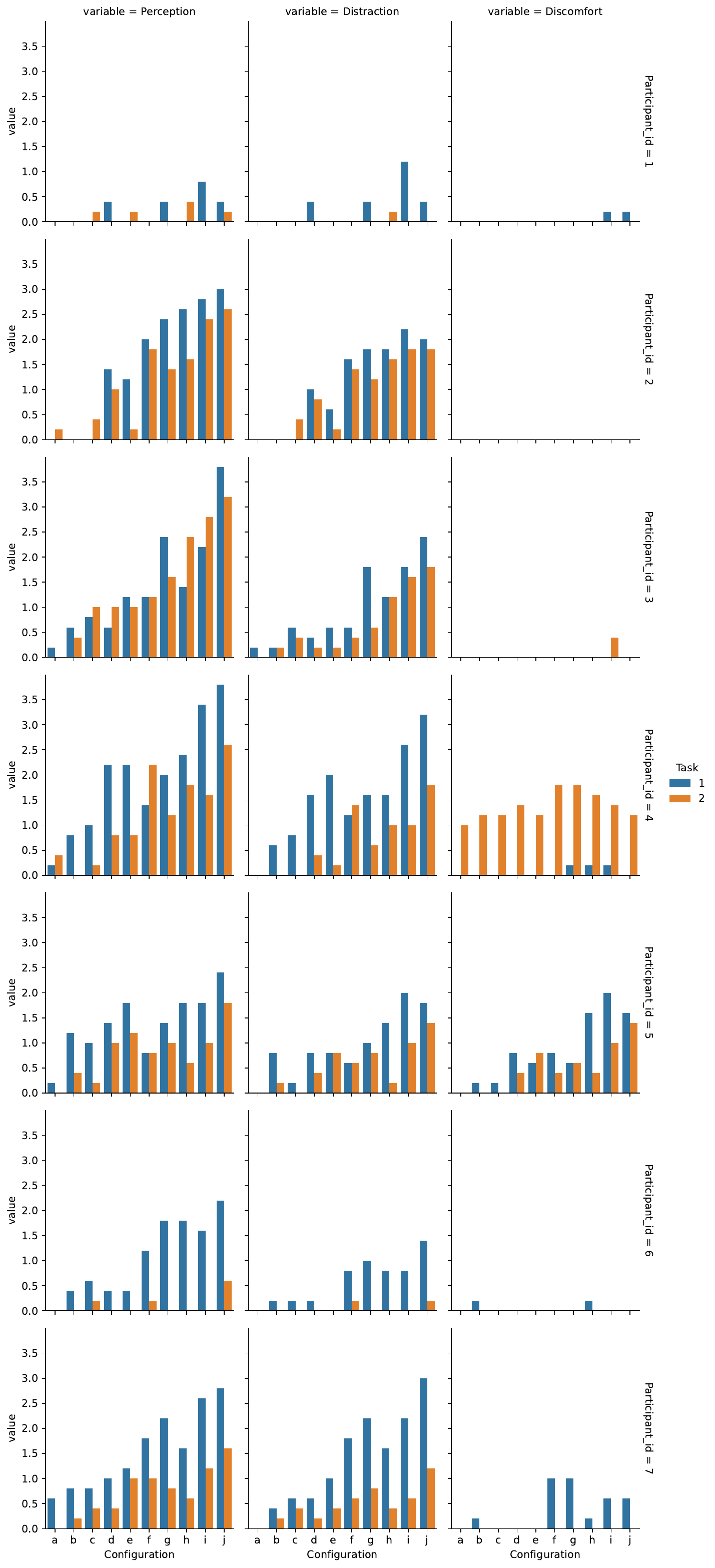}  & \\
\end{tabular}
  \caption{Likert scale user feedback per participant (rows, wrapped at participant 7) per configuration (x-axes) per task (bar color), averaged over backgrounds and shapes configurations.}
  \Description{15 rows of bar charts over two columns, one row per participant, faceted horizontally by perception, distractions, and discomfort. The local y axis depicts the Likert-scale responses, the local x axis enumerates the 10 frame rate configurations. Each configuration shows a bar for task 1 next to a bar for task 2.}
  \label{fig:pdd_part_task}
\end{figure*}


\begin{figure*}[ht!]
  \centering
  \includegraphics[width=.99\linewidth]{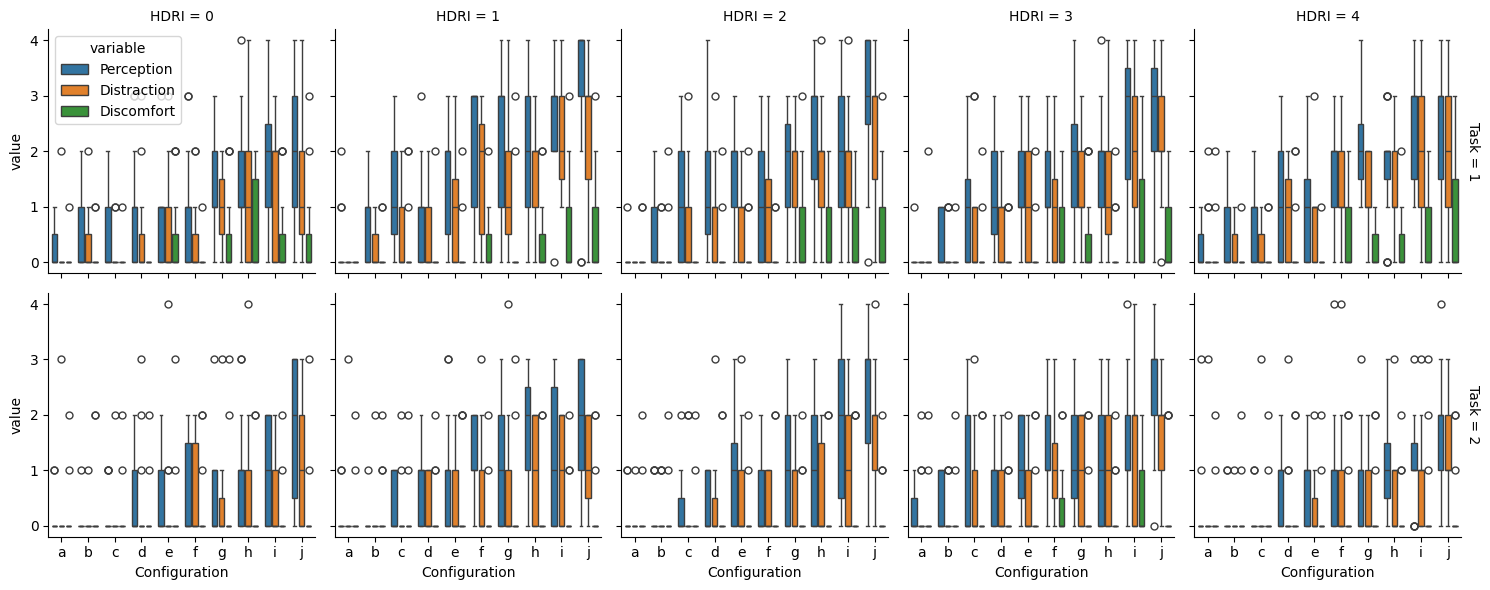}  
  \caption{Boxplots showing results of user feedback grouped by scene environment. Top: Gaze follow task. Bottom: Gaze center task.
  }
  \Description{Boxplots faceted horizontally over the environment and vertically over the tasks. Local x axes depict the frame rate configuration, local y axes the Likert-scale responses. Each configuration shows a boxplot for perception, distraction, and discomfort values each.}
  \label{fig:results_environments}
\end{figure*}

\begin{figure*}[ht!]
  \centering
  \includegraphics[width=.5\linewidth]{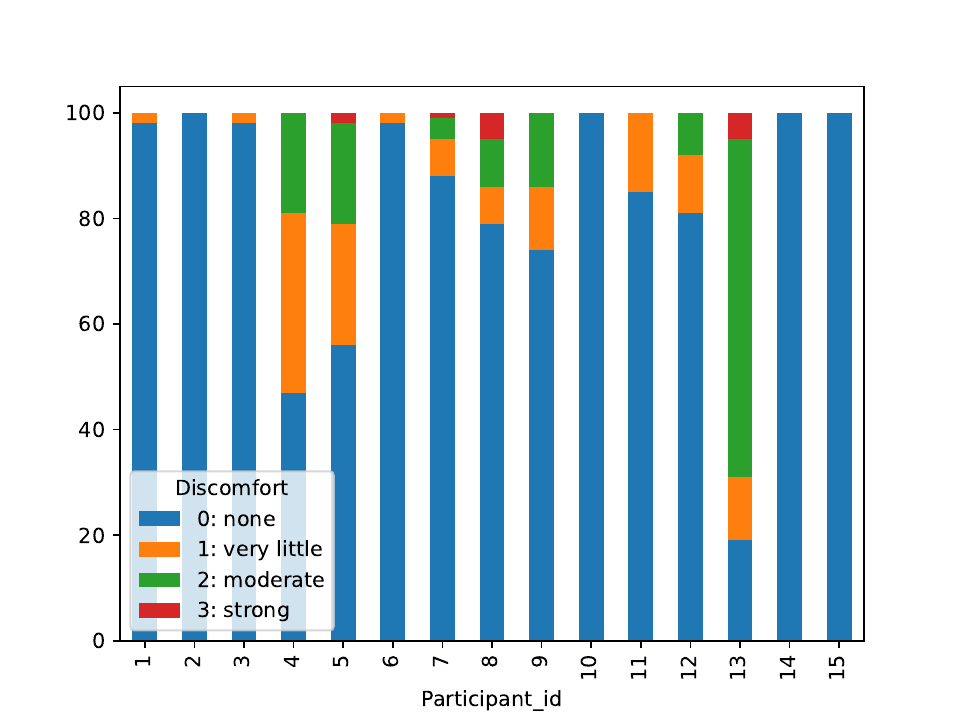}  

  \caption{Discomfort rating occurrences per participant over all rounds.}
  \Description{Stacked bar chart over all occurring discomfort values of all participant ids depicted on the x axis. Most the the chart shows the color for "none", and only three participants show obvious signs (more than a third) of very little or more discomfort.}
  \label{fig:discomfort_per_participant}
\end{figure*}


\begin{figure*}[!htb]
    \centering
    \includegraphics[width=0.25\linewidth]{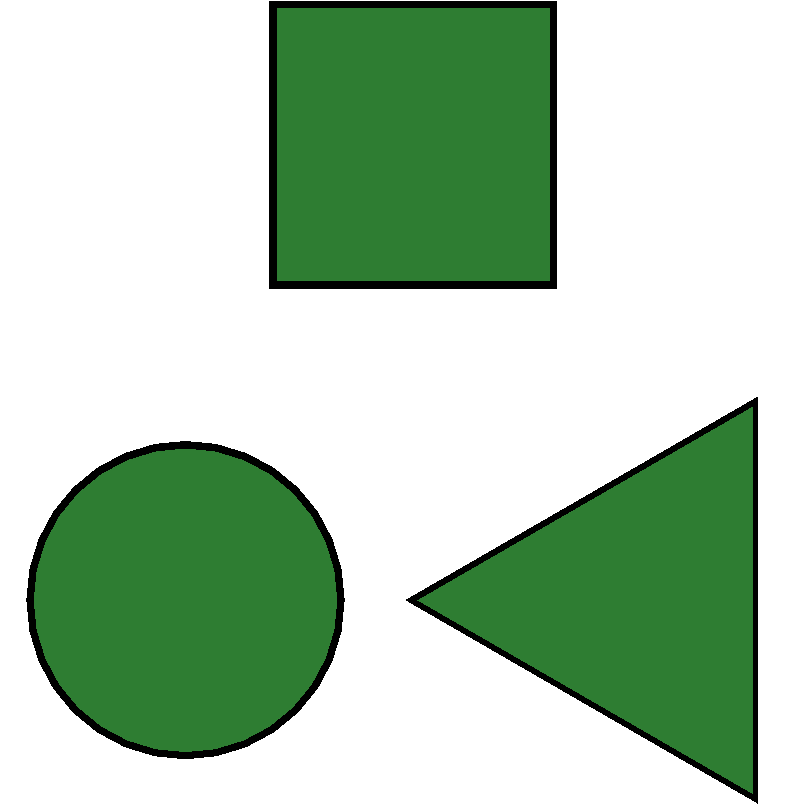}
    \includegraphics[width=0.25\linewidth]{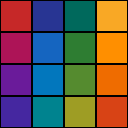}
    \caption{The geometry of the moving objects (left) and the color they can be (right). A Black border is added to ensure distinguishability of the objects even when overlapping.}
    \Description{Two subfigures. On the left, the three moving shapes are shown: a square, a circle, and a triangle. On the right, a 4 by 4 matrix shows all possible colors that could be assigned to the shapes.}
    \label{fig:object_col_geo}
\end{figure*}


\begin{figure*}[ht!]
\centering
\subfloat[\label{fig:conf_a}$100\%$] {\includegraphics[width=.24\linewidth]{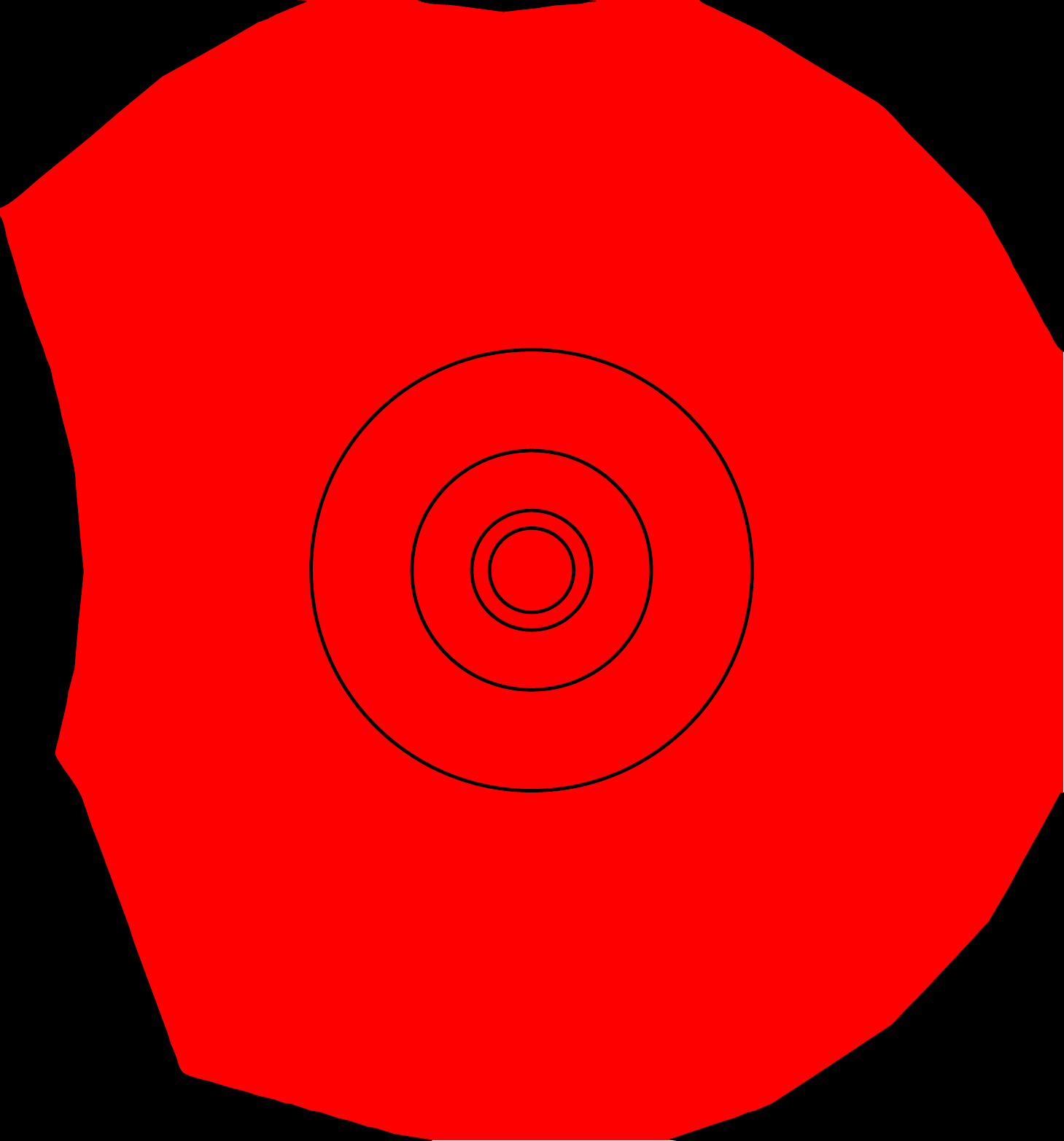}}\hfill
\subfloat[\label{fig:conf_b}$57,6\%$]{\includegraphics[width=.24\linewidth]{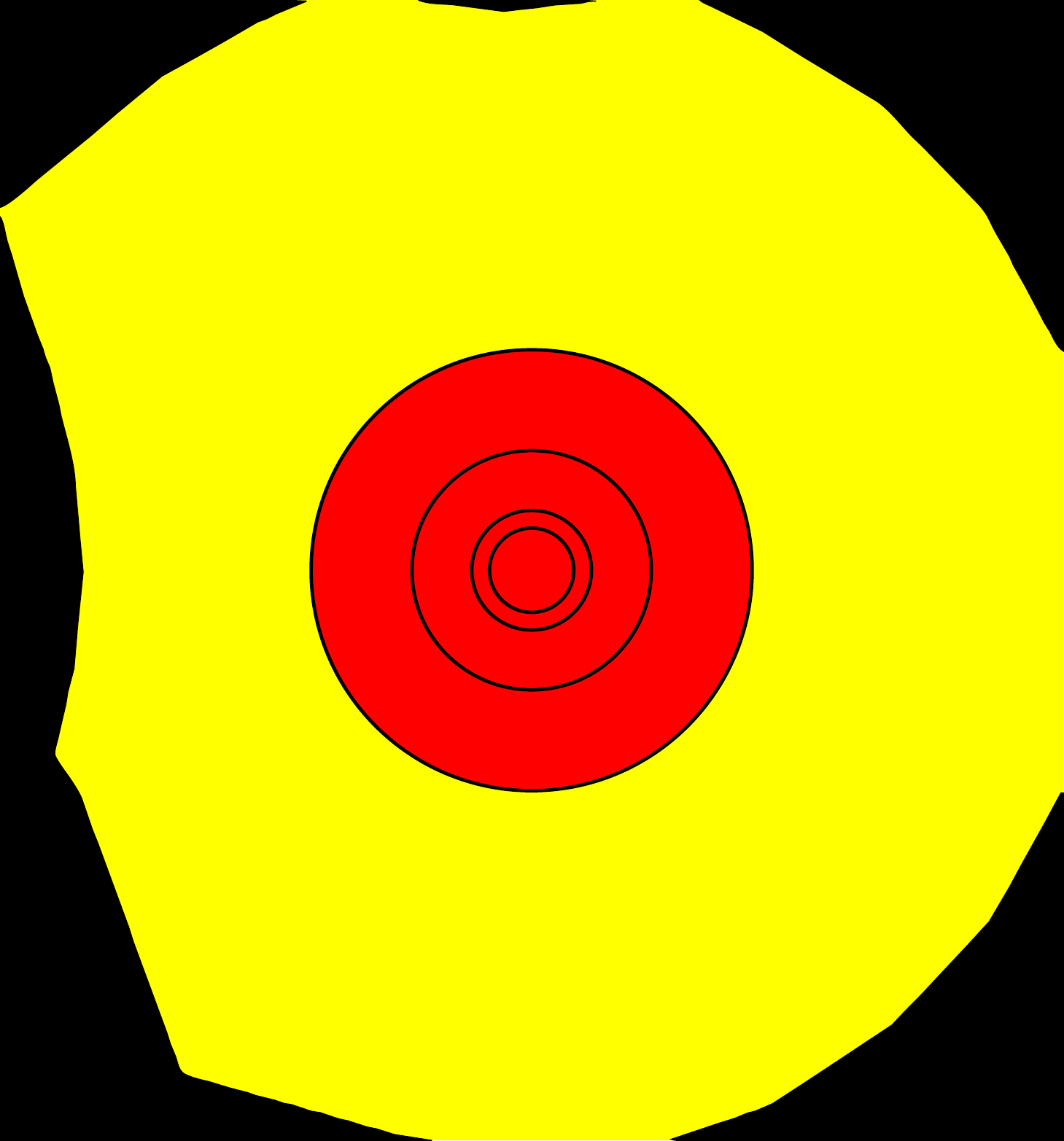}}\hfill
\subfloat[\label{fig:conf_c}$52,2\%$]{\includegraphics[width=.24\linewidth]{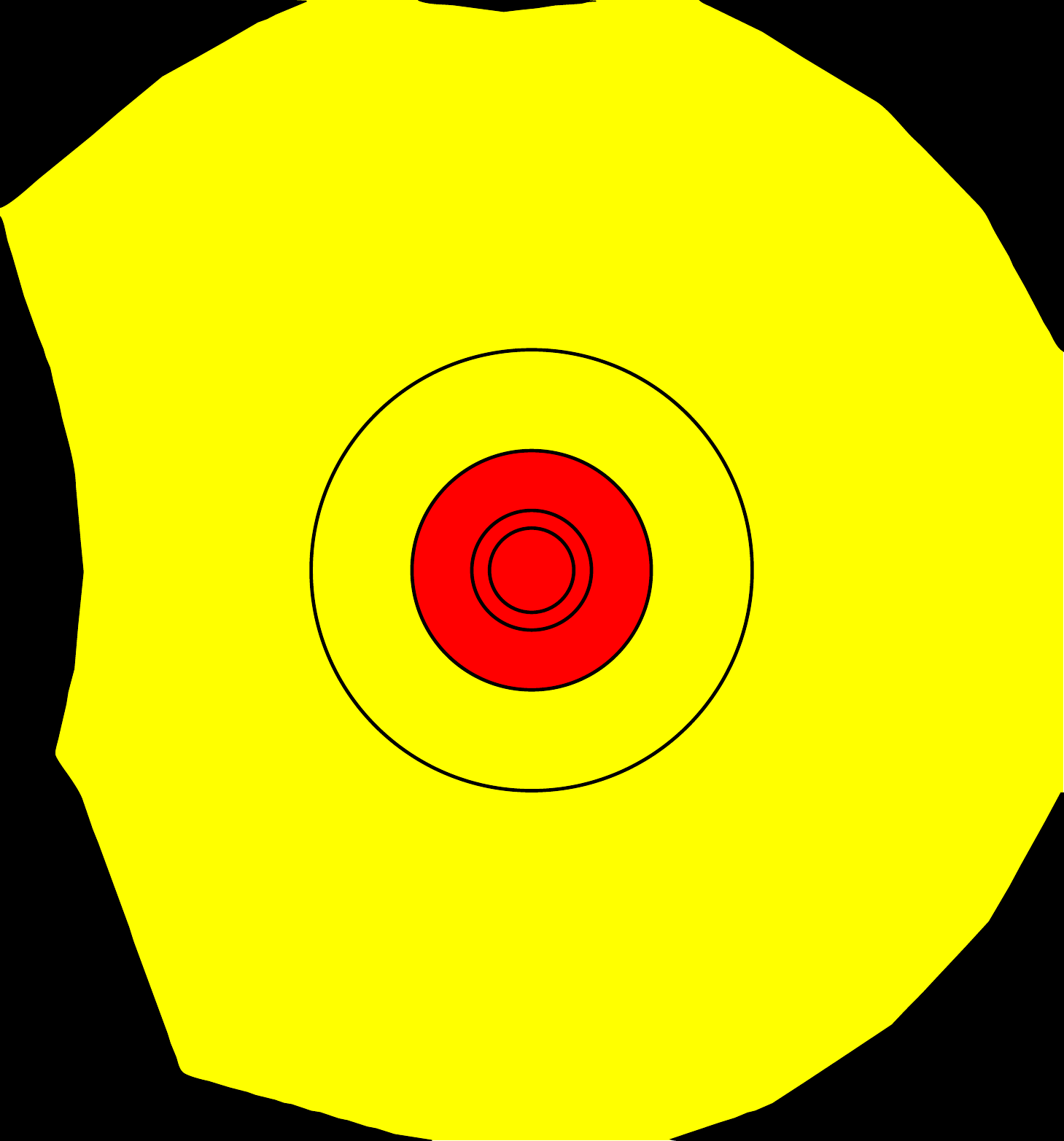}}\hfill
\subfloat[\label{fig:conf_d}$50,2\%$]{\includegraphics[width=.24\linewidth]{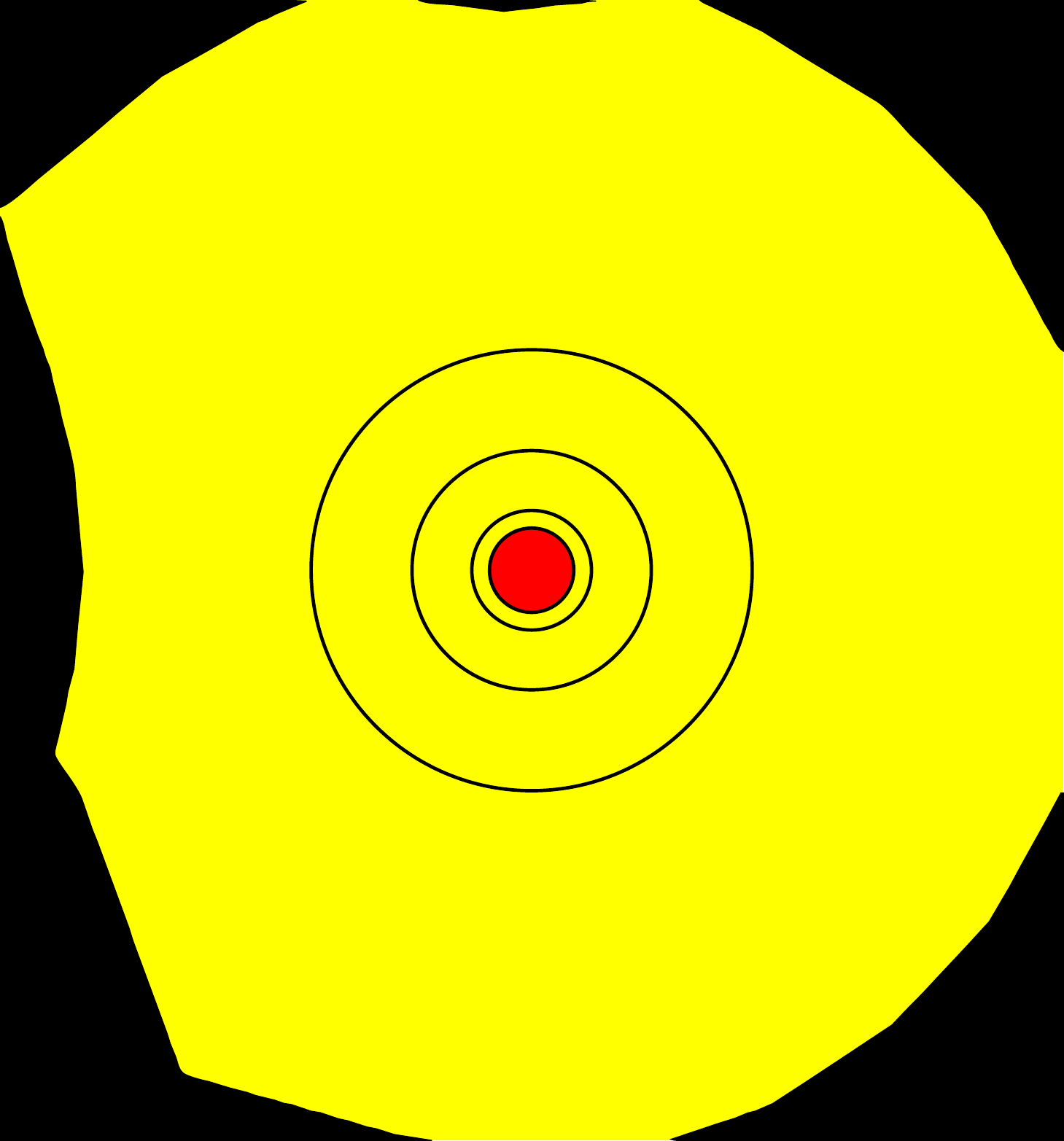}}\hfill\\
\subfloat[\label{fig:conf_e}$36,4\%$]{\includegraphics[width=.24\linewidth]{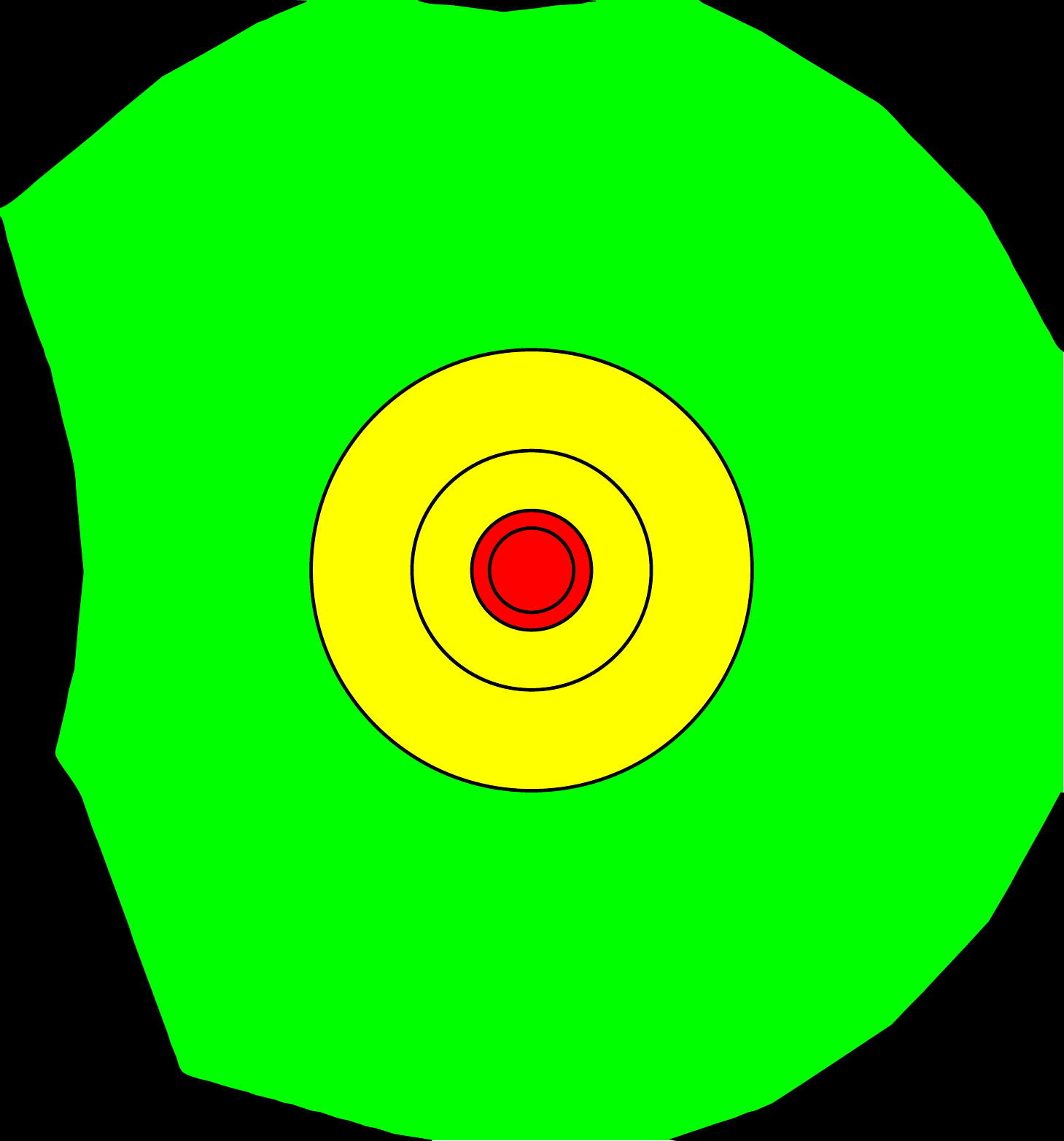}}\hfill
\subfloat[\label{fig:conf_f}$34,6\%$]{\includegraphics[width=.24\linewidth]{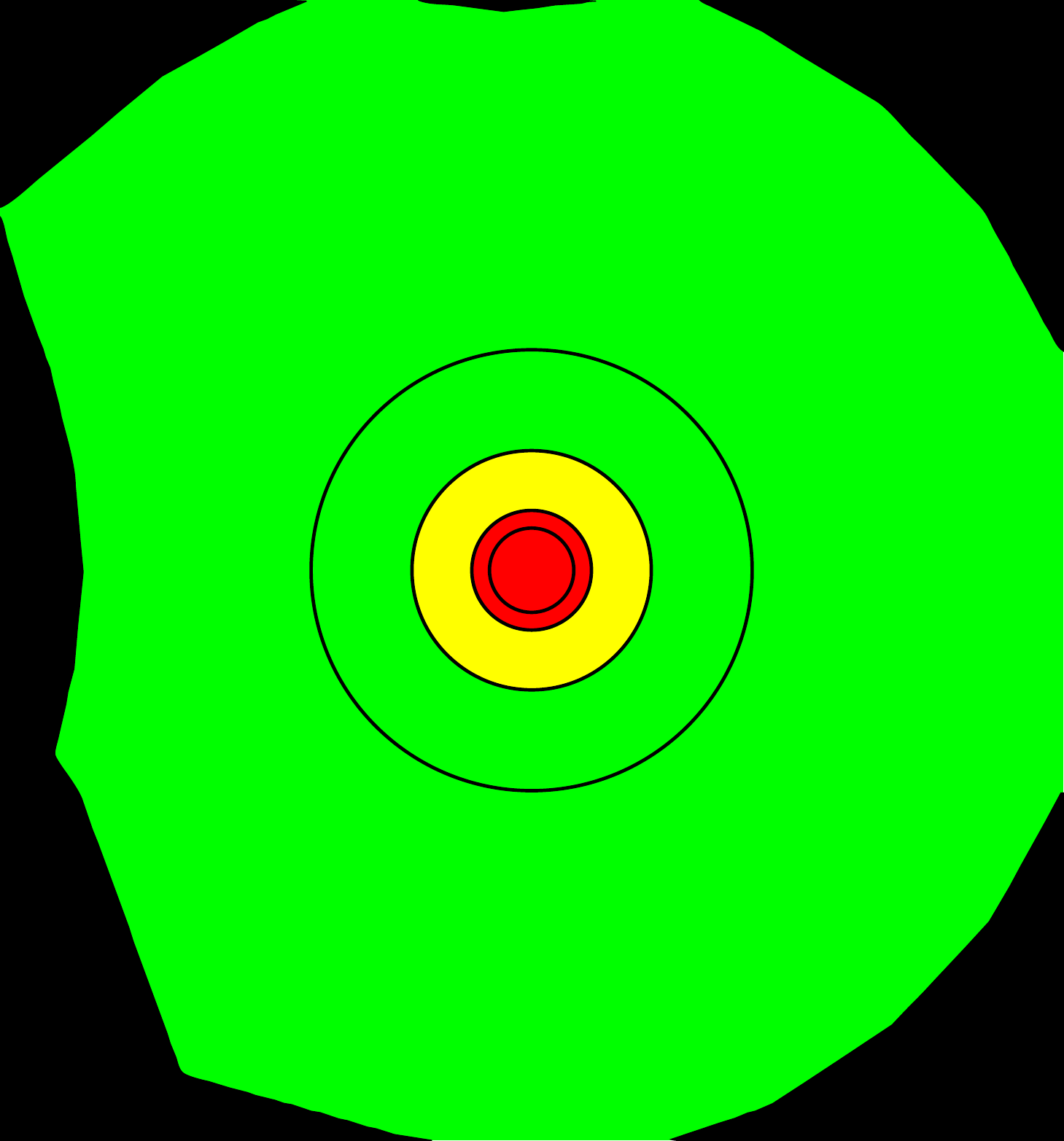}}\hfill
\subfloat[\label{fig:conf_g}$34,4\%$]{\includegraphics[width=.24\linewidth]{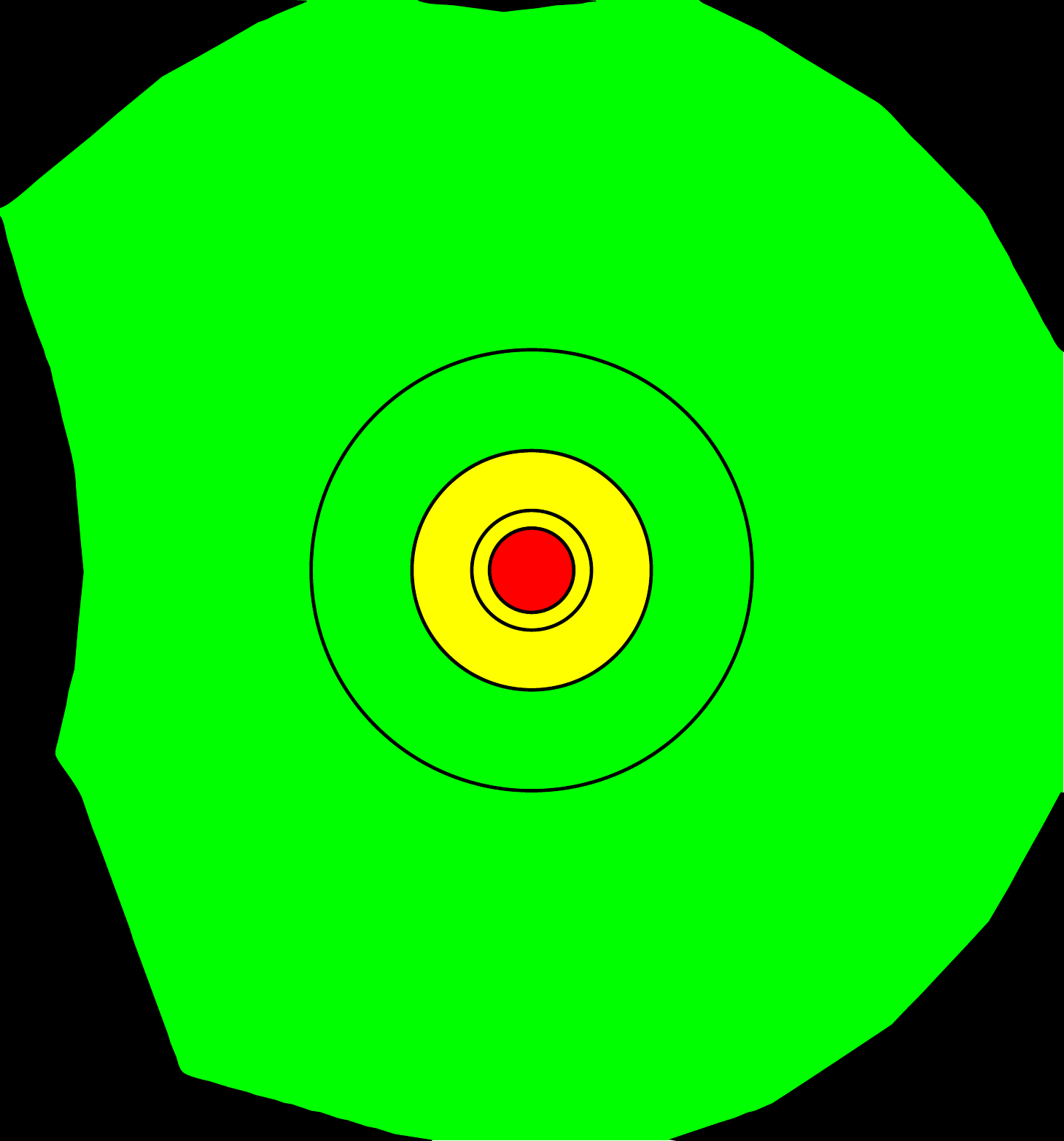}}\hfill
\subfloat[\label{fig:conf_h}$27,3\%$]{\includegraphics[width=.24\linewidth]{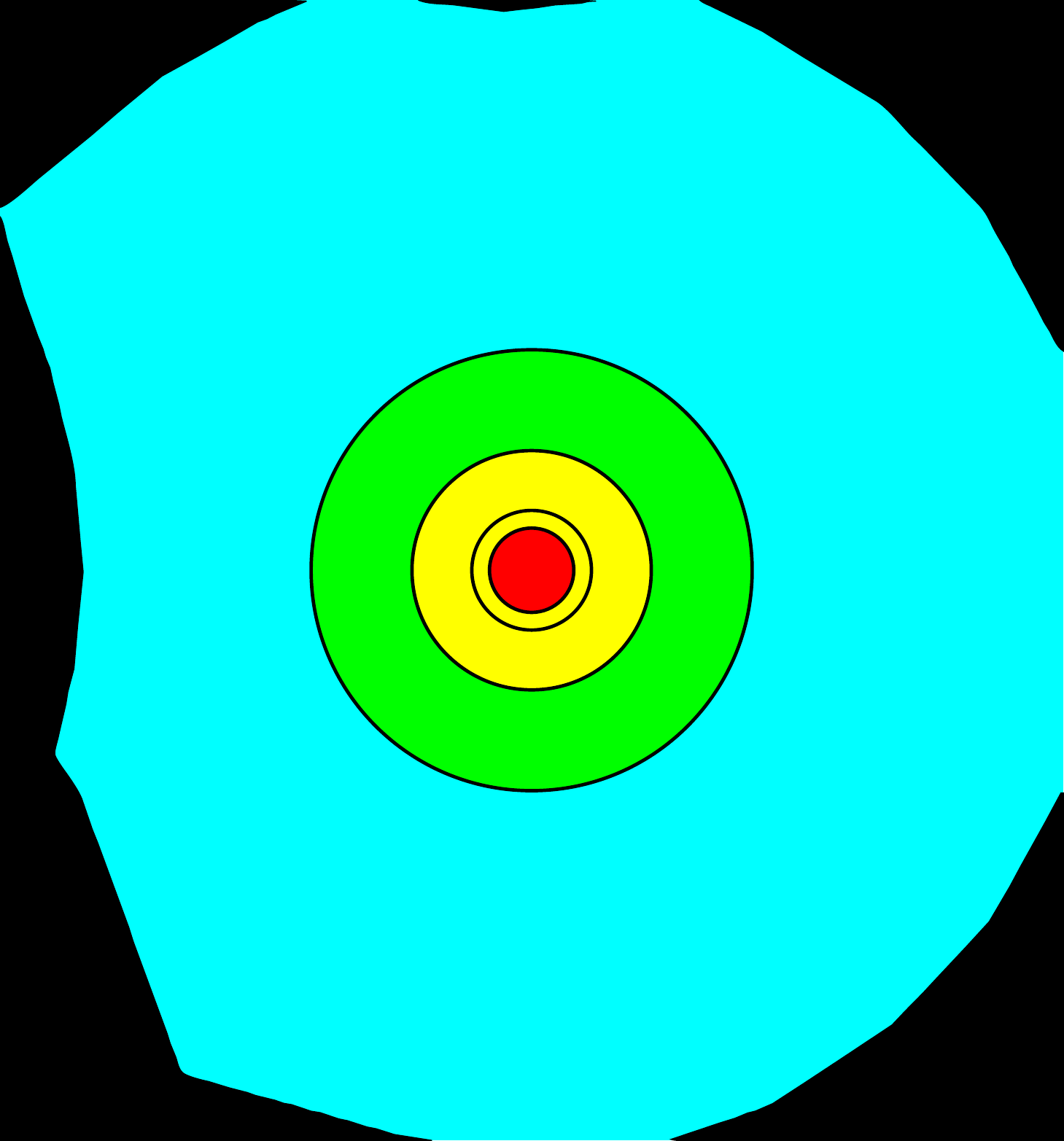}}\hfill\\
\subfloat[\label{fig:conf_i}$26,7\%$]{\includegraphics[width=.24\linewidth]{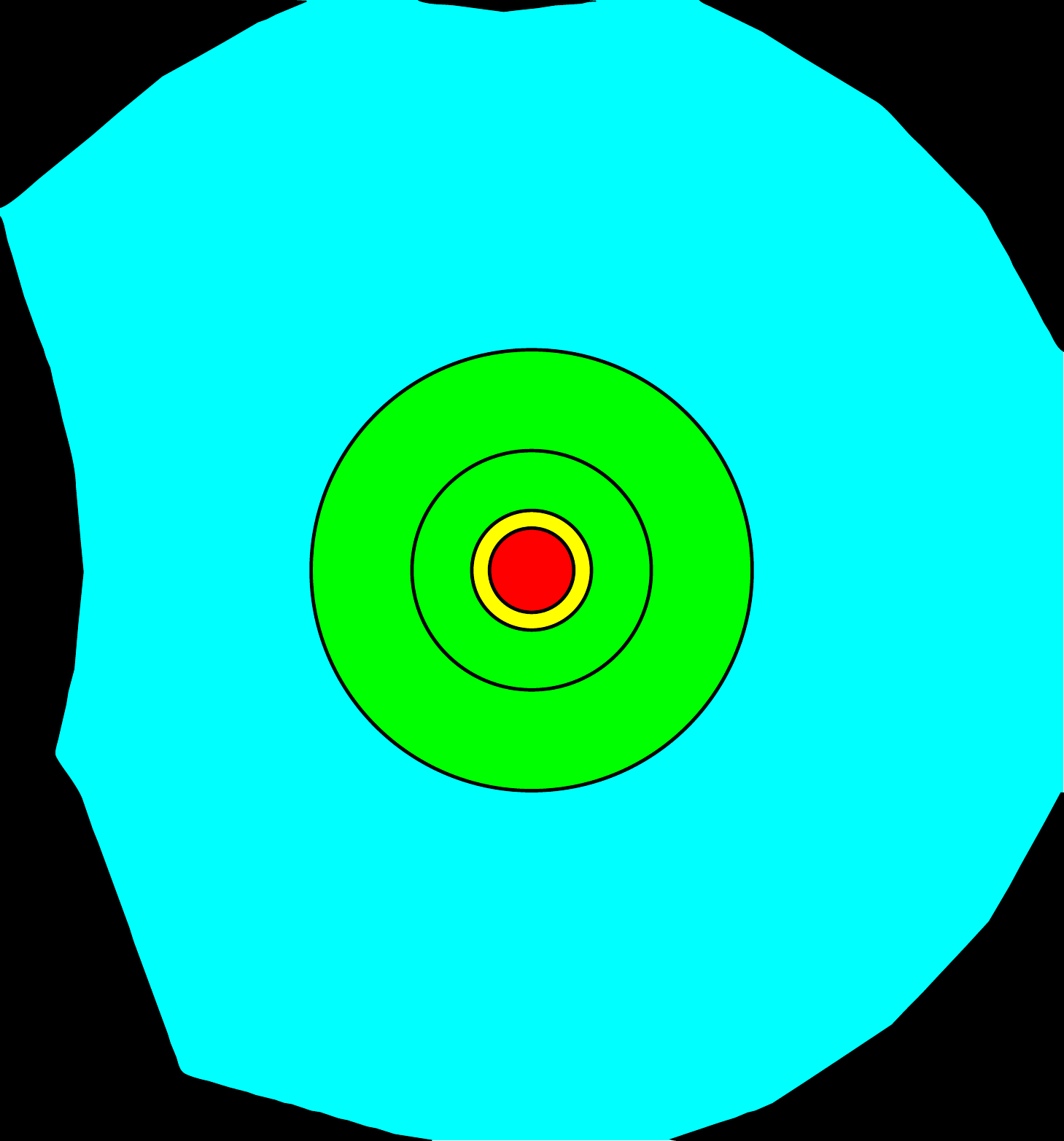}}\hfill
\subfloat[\label{fig:conf_j}$21,5\%$]{\includegraphics[width=.24\linewidth]{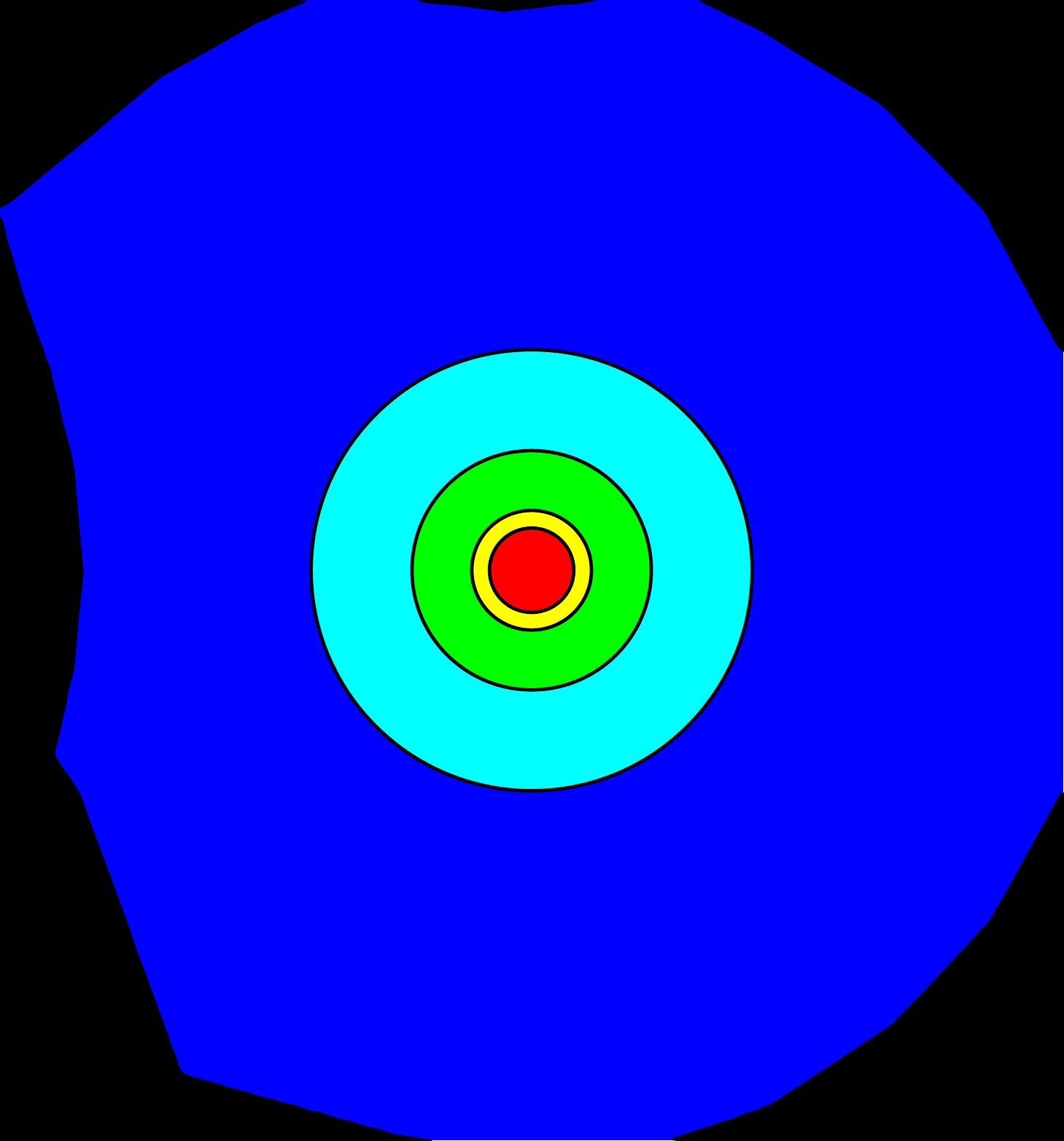}}\hfill
\subfloat[\label{fig:conf_angle}Max. eccentricity]{\includegraphics[width=.24\linewidth]{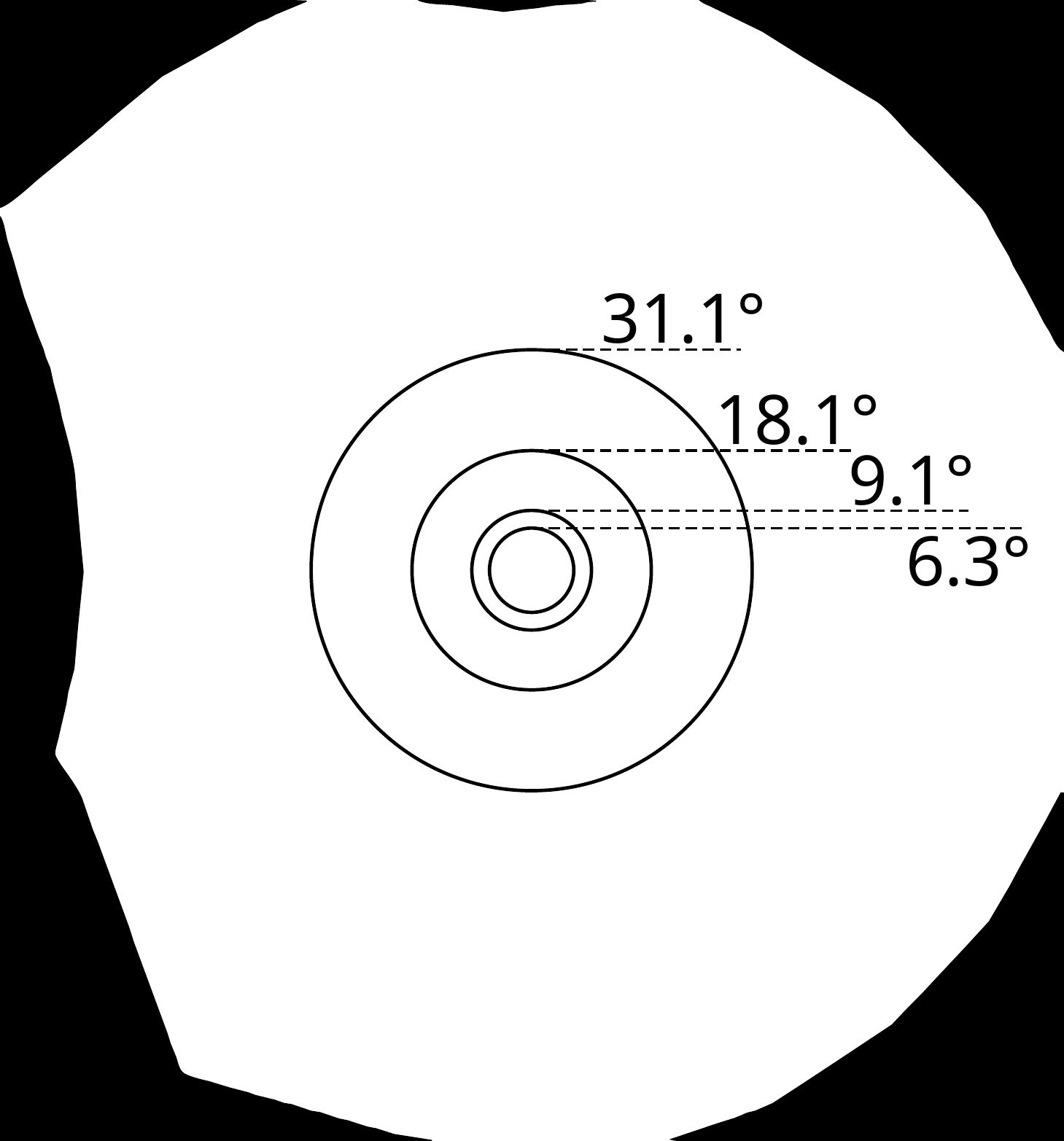}}\hfill
\subfloat[\label{fig:conf_key}Reduction scheme]{\includegraphics[width=.24\linewidth]{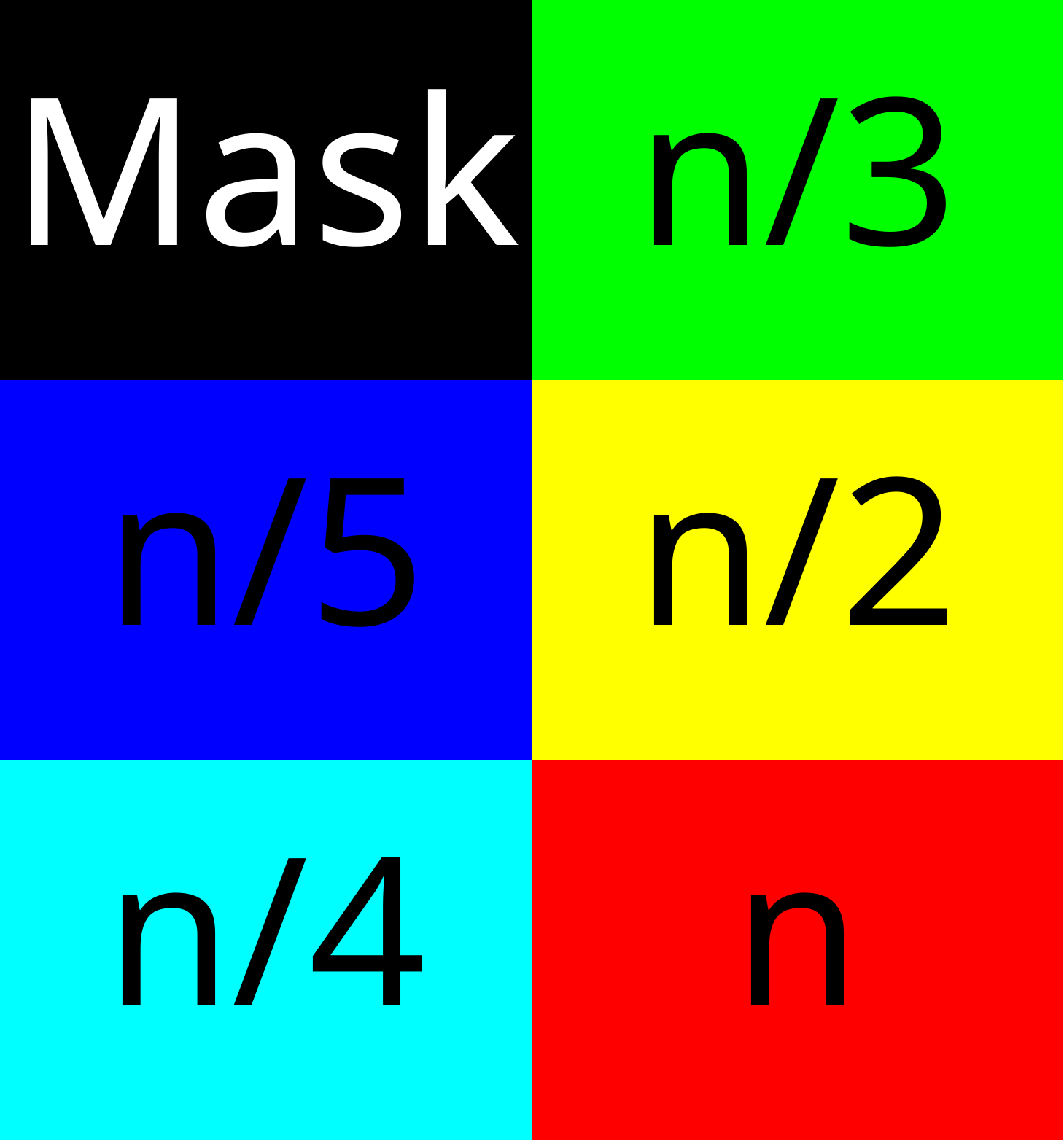}}\hfill
\caption{Visualization of the mask for each framework configurations used in the user study, with the user's gaze focused at the center of the display. The corresponding pixel reduction rate in \%. The colors of the regions represent the effective frame rate for the region. The maximum degree of eccentricity for each region is given in (k).}
\Description{A 4 by 3 image matrix, showing first the 10 different frame rate configurations with each concentric segment colored by the fraction of frame refreshes occurring (every frame up to every fifth frame). The last two pictures contain the concentric segments with their respective degree of eccentricity and the legend mapping frame rates to colors.}
\label{fig:all_config}
\end{figure*}


\begin{figure*}[ht!]
  \centering
  \includegraphics[width=.24\linewidth]{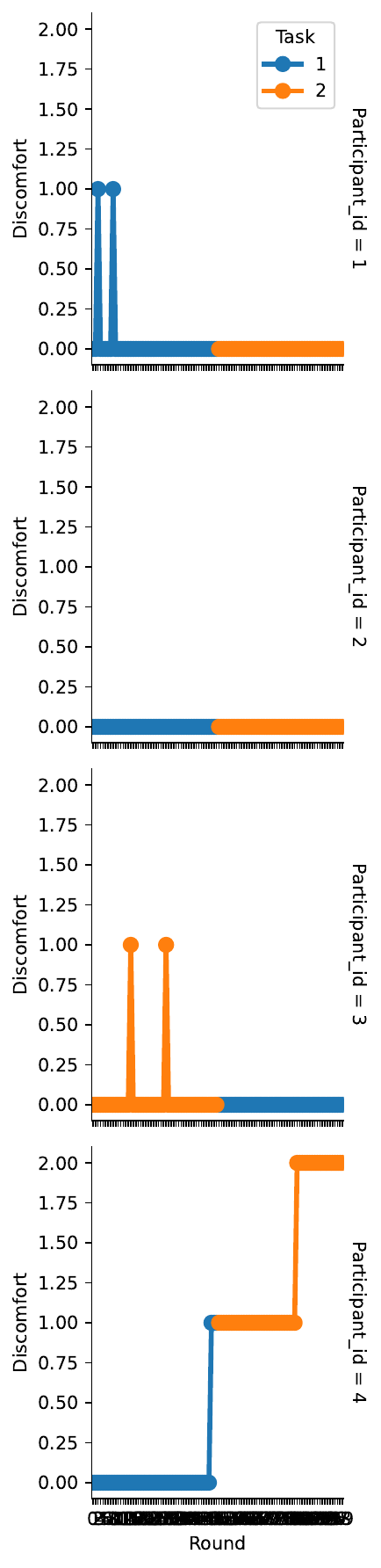}  
  \includegraphics[width=.24\linewidth]{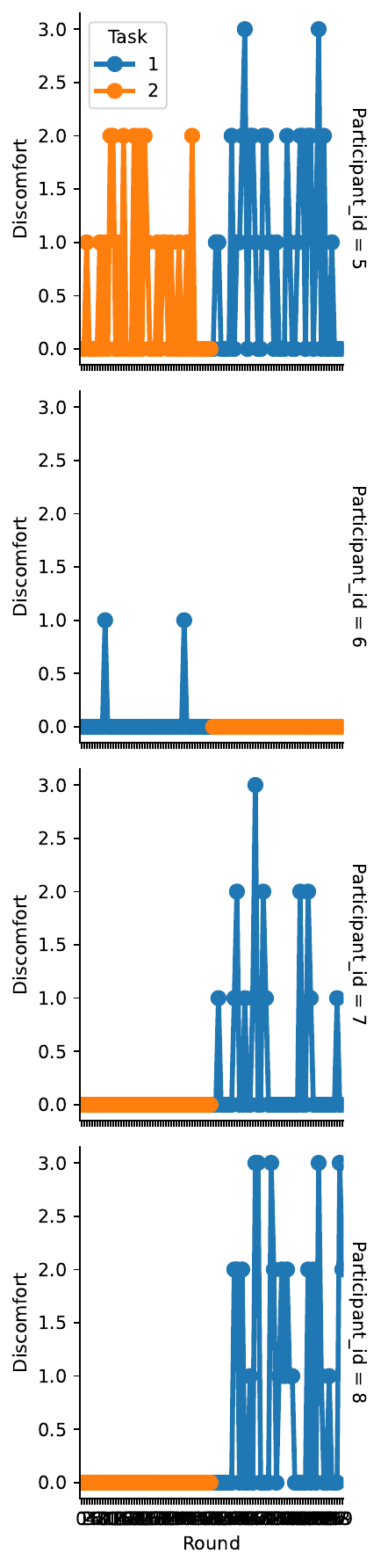}  
  \includegraphics[width=.24\linewidth]{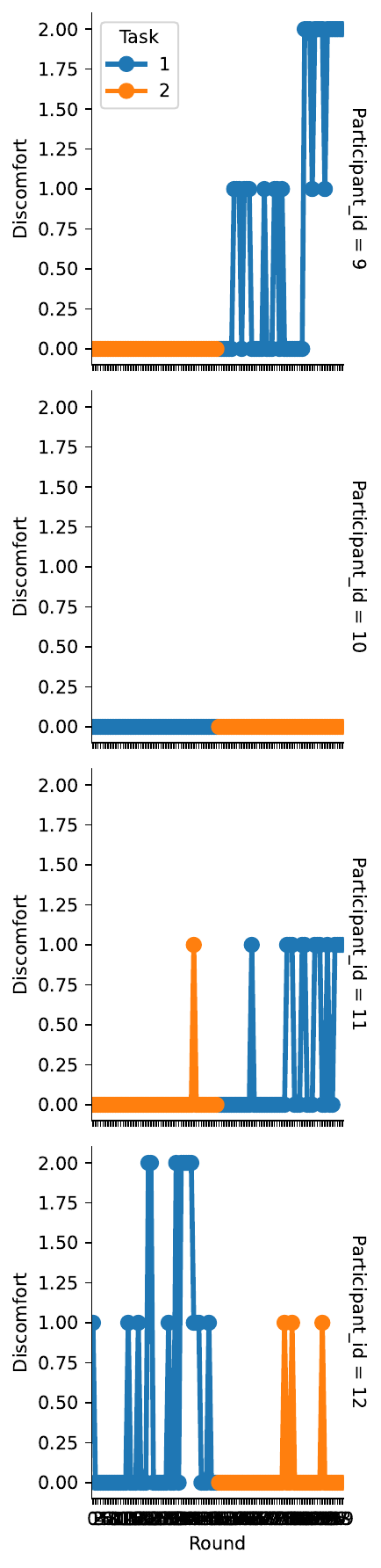}  
  \includegraphics[width=.24\linewidth]{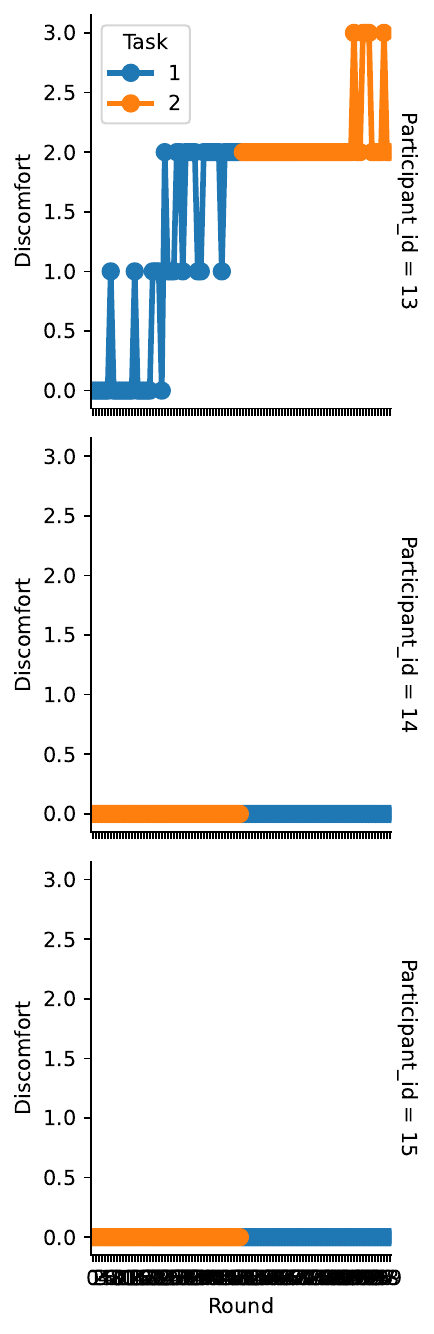}  

  \caption{Discomfort over rounds (x-axes) for each participant (rows), colored by task. Rows are wrapped at participants 4, 8, and 12. Note how only participants 4 and 13 have a clearly increasing discomfort over time.}
  \Description{A matrix of 4 by 4 line plots showing discomfort over trial time for each participant. Lines are colored according to the task: 7 participants started with task 1. One matrix element is empty since there are only 15 participants.}
  \label{fig:discomfort_over_time}
\end{figure*}

\end{document}